\documentclass[journal,twoside,web]{ieeecolor}
\usepackage{generic}
\usepackage{cite}
\usepackage{amsmath,amssymb,amsfonts}
\usepackage{algorithmic}
\usepackage{graphicx}
\usepackage{algorithm,algorithmic}
\usepackage{hyperref}
\hypersetup{hidelinks=true}
\usepackage{textcomp}
\usepackage{epstopdf}

\def\BibTeX{{\rm B\kern-.05em{\sc i\kern-.025em b}\kern-.08em
    T\kern-.1667em\lower.7ex\hbox{E}\kern-.125emX}}
\begin{document}
\title{Destination-Constrained Linear Dynamical System Modeling in Set-Valued Frameworks}
\author{Xiaowei Yang, Haiqi Liu, Fanqin Meng and Xiaojing Shen 
\thanks{This work was supported in part by  Sichuan Youth Science and Technology Innovation Team (Grant Nos. 2022JDTD0014, 2021JDJQ0036).  (Corresponding author: Haiqi Liu.)}
\thanks{Xiaowei Yang, Haiqi Liu, and Xiaojing Shen are with the College of Mathematics, Sichuan University, Chengdu, Sichuan 610064, China (e-mail:yxw8290@163.com; haiqiliu0330@163.com; shenxj@scu.edu.cn). }
\thanks{Fanqin Meng is with the Artificial Intelligence Key Laboratory of Sichuan Province, Yibin, Sichuan 610064, China (e-mail: mengfanqin2008@163.com). }}

\maketitle

\begin{abstract}
Directional motion towards a specified destination is a common occurrence in physical processes and human societal activities. Utilizing this prior information can significantly improve the control and predictive performance of system models. This paper primarily focuses on reconstructing linear dynamic system models based on destination constraints in the set-valued framework. We treat destination constraints as inherent information in the state evolution process and employ convex optimization techniques to construct a coherent and robust state model. This refined model effectively captures the impact of destination constraints on the state evolution at each time step. Furthermore, we design an optimal weight matrix for the reconstructed model to ensure smoother and more natural trajectories of state evolution. We also analyze the theoretical guarantee of optimality for this weight matrix and the properties of the reconstructed model. Finally, simulation experiments verify that the reconstructed model has significant advantages over the unconstrained and unoptimized weighted models and constrains the evolution of state trajectories with different starting and ending points.
\end{abstract}

\begin{IEEEkeywords}
Set-Valued, Destination Constraints, Modeling, Linear Dynamic Systems
\end{IEEEkeywords}

\section{Introduction}
\IEEEPARstart{I}{n} recent years, there has been a significant surge in research focused on modeling dynamic systems in various scenarios, as outlined in \cite{c1} and \cite{c2}. Dynamic system modeling and prediction play a crucial role in a wide range of fields, including air traffic management \cite{c3,c3-1,c3-2}, target tracking \cite{c4,c4-1,c4-2}, and robotics technology \cite{c5,c5-1,c5-2}. Many practical applications involve dynamic systems operating under constraint conditions, where state variables must adhere to specific physical principles or mathematical relationships. These constraint conditions contain valuable information about the intrinsic properties of dynamic systems, and effectively leveraging these constraints can enhance the accuracy and robustness of state estimation \cite{c6,c7,c8}. In many dynamic systems, state trajectories are influenced by equality constraints \cite{c9}. Several key results have been achieved in the context of linear equation-constrained state estimation. For instance, dimensionality reduction methods have proven effective in transforming constrained state estimation problems into dimensionality-reduced unconstrained problems \cite{c10}. Another popular approach involves projection methods, which utilize constrained optimization techniques to project unconstrained estimates onto the constraint subspace \cite{c11,c12}. Pseudo-measurement methods have also been applied to equality-constrained state estimation, treating constraints as pseudo-measurements and thereby transforming state estimation problems into standard filtering problems with two types of measurements \cite{c13,c14,c15}. State trajectories can be mathematically modeled using appropriate stochastic processes. Trajectory modeling with destination information and intent inference has diverse applications in fields such as air/ground traffic \cite{c16,c17}, missile systems \cite{c18}, and human-machine interaction \cite{c19}. Various methods for trajectory modeling, prediction, and intent inference have been proposed \cite{c20,c21,c22}. 

Departures, waypoints, and destinations contain valuable information for system modeling \cite{c23}. Researchers have strived to improve tracking performance by leveraging destination information, with the initial concept introduced by \cite{c24} focusing on maritime surveillance. Building on this work, researchers like \cite{c25} and \cite{c26} have conducted systematic research, emphasizing modeling, estimation, and smoothing for reciprocal processes. Recent research advances have highlighted the utility of Conditional Markov (CM) processes as a more natural and powerful generalization over traditional Markov processes. This insight has driven research by \cite{c27} and \cite{c28}, which comprehensively explore CM sequence modeling and establish state models with destination information. Additionally, literature such as \cite{c29} and \cite{c30} delves into constructing linear dynamic models with destination constraints and addressing state estimation problems under the assumption of Gaussian white noise.

In some practical applications, obtaining precise statistical properties of noise is often challenging due to limitations in manpower and resources \cite{c31}. This inherent uncertainty significantly impacts the performance of state models. Thus, it is practical to account for the uncertainty of unknown noise by defining bounded regions, as discussed in \cite{c32,c33,c34,c35}. In the setting of  unknown but bounded noise, set-valued filters were introduced in the late 1960s by \cite{c36}. These filters, unlike Bayesian methods, aim to provide assured bounds for the system's state while handling uncertain effects of noise with unknown but limited magnitude, without making assumptions about its stochastic characteristics \cite{c37}. Subsequently, these filters have received considerable attention and rigorous research from numerous scholars in dynamic systems characterized by unknown but bounded noise \cite{c38,c39,c40}. It's worth noting that most research on set-valued filters has primarily focused on the state estimation problem of unconstrained linear dynamic systems \cite{c41,c42,c43,c44}. Early literature on the estimation problem with state constraints includes \cite{c45} and \cite{c46}. Recently, some scholars have investigated the incorporation of different constraint conditions into target state updates in the set-valued framework, proposing set-valued filters and algorithms for constrained state estimation in the presence of unknown but bounded noise, yielding promising results \cite{c47,c48}. These studies mainly address set-valued filtering for constrained dynamic systems with unknown but bounded noise, contributing to improved reliability and robustness of state models in engineering applications.

From the comprehensive literature review above, it can be observed that a substantial amount of research and development has been conducted on system models with destination constraints within a stochastic framework. However, to the best of our knowledge, there is still a relatively underexplored area regarding modeling dynamic systems with destination constraints in the set-valued framework. In this paper, we concentrate on a general, systematic, and mathematically sound modeling of dynamic systems with destination constraints in the set-valued framework. Furthermore, we delve into an in-depth analysis of the theoretical properties of the constructed model. The key contributions of this paper can be summarized as follows:
\begin{itemize}
	\item \textit{Destination-constrained model reconstruction.} Drawing on convex optimization techniques, in scenarios where noise is unknown but bounded, we reconstruct a dynamic system model with destination constraints. Departing from the view of destination constraints as mere additional restrictions on the final state, we recognize them as intrinsic properties throughout the state evolution process. State projection methods are employed to seamlessly integrate destination constraints into the original dynamical model, resulting in a consistent closed-form model that exhibits superior performance compared to unconstrained dynamic system models in the set-valued framework. Additionally, this modeling approach is not limited by the distribution of noise.
\end{itemize}
\begin{itemize}
	\item \textit{Optimal weight matrix design.} We design an optimal weight matrix in the set-valued framework to enhance the smoothness and naturalness of state evolution trajectories in the reconstructed model. To address the challenge of unknown but bounded process noise, a set of bounded ellipsoids is employed to cover all possible noise scenarios. The paper leverages the affine properties of ellipsoids, ensuring that ellipsoids covering the original states remain lossless throughout the computation process.
\end{itemize}
\begin{itemize}
	\item \textit{Exploration of theoretical properties.} The paper establishes the optimal theoretical properties of the weight matrix for the reconstructed model in the set-valued framework. The relationship between the magnitude of process noise generated by the reconfigured state model with an optimal weight matrix and the unconstrained state model is investigated at corresponding time steps. Additionally, the study examines how the process noise level changes as the moving target approaches the destination under mild conditions.
\end{itemize}
\begin{itemize}
	\item \textit{Validation through simulation experiments.} Extensive simulation experiments have been conducted to validate the proposed dynamic system model with destination constraints in set-valued scenarios. Comprehensive numerical simulations demonstrate the superior performance and stability of the proposed model compared to both the unconstrained state model and the state model without weight matrix optimization in the set-valued context. The experiments also illustrate the impact of destination constraints on the system's state evolution process.
\end{itemize}

Briefly, this paper introduces innovative approaches in the application of convex optimization, weight matrix design, and theoretical exploration in the set-valued framework for modeling linear dynamical systems with destination constraints. The presented simulation experiments provide empirical support for the effectiveness and robustness of the proposed methods.

The remainder of the paper is organized as follows. In Section II, we address the problem of modeling linear dynamic systems with destination constraints under the setting of unknown but bounded process noise. Section III applies convex optimization techniques in the set-valued framework to reconstruct dynamic system models with destination constraints. To ensure that the reconstructed model generates state trajectories that better adhere to real-world evolution patterns, Section IV delves into the design and the optimality theory of the weight matrix for the reconstructed model under the set-valued framework. Section V discusses the relevant  properties of the reconstructed model with the optimal weight matrix in the set-valued context. In Section VI, simulation experiments are conducted to showcase the superiority and effectiveness of the reconstructed model. Section VII provides the conclusion. Finally, the Appendix offers detailed proofs for all propositions presented in this paper.

The majority of the notation employed in this paper conforms to standard conventions. The $n$-dimensional real-valued vector space is denoted as $\mathbb{R}^{n} $, while the $n\times m$-dimensional real-valued matrix space is denoted as $\mathbb{R}^{m\times n} $. When provided with a matrix $Z$, the notations $Z^{-1}$, $Z^{T}$, $\text{tr}(Z)$ and $\text{rank}(Z)$ denote the matrix's inverse, transpose, trace, and rank, respectively. For a positive definite (positive semi-definite) matrix $Z$ denoted as $Z\succ \mathbf{0}$ ($Z \succeq \mathbf{0}$), where the zero matrix $\mathbf{0}$ has the appropriate dimensions.

\section{Problem Description}
Motion behaviors or phenomena with destination orientation are quite common in actual motion scenarios. For example, consider a time-varying discrete dynamic system that satisfies the following destination constraint:
\begin{align}\label{eq:1}
	x_N \in\mathcal{G},
\end{align}
where $x_{N}\in\mathbb{R}^{n}$ represents the motion state of the target when it reaches the destination at time step $N$, and $\mathcal{G}$ is a known region (i.e., prior information).
To initiate our exploration, we introduce some basic concepts in the ellipsoidal set-valued context.

A set that satisfies the following form
\begin{align}\label{eq:2}
	\mathcal{E}=\{x\in\mathbb{R}^{n}|(x-\hat{x})^{T}P^{-1}(x-\hat{x})\le1\}
\end{align}
is called an ellipsoid. Here, $\hat{x} $ and $P$ denote the center and shape matrix of the ellipsoid, respectively, and $P$ is a positive definite matrix. The ellipsoid set can also be equivalently written as
\begin{align}\label{eq:3}
	\mathcal{E}=\{x\in\mathbb{R}^{n}|x=\hat{x}+Er,\|r\|\le1\},
\end{align}
where $E$ is the Cholesky decomposition of $P$, i.e., $P=EE^{T}$, and $\|\cdot\|$ denotes the Euclidean norm of a vector. It can be seen from \eqref{eq:3} that the ellipsoid is obtained by affine transformation of the unit ball. Furthermore, it's important to note that an affine transformation of an ellipsoid does not alter its shape, making it conformal \cite{c50}. This can be expressed as follows: 
\begin{align*}
	\tilde{\mathcal{E} } & = U\mathcal{E}+b  = \{z\in\mathbb{R}^{n}|z = Ux+b,x\in \mathcal{E}\}\\
	& = \{z\in\mathbb{R}^{n}|(z-U\hat{x}-b)^{T}(UPU^{T})^{-1}(z-U\hat{x}-b)\le 1\},
\end{align*}
in which $U$ and $b$ denote a matrice and a vector with appropriate dimensions, respectively.
Clearly, after an affine transformation, $\tilde{\mathcal{E}}$ remains an ellipsoid with its center at $U\hat{x}+b$ and shape matrix $UPU^{T}$. 

Let us consider a linear dynamic system with unknown but bounded noises as follows:
\begin{align}\label{eq:4}
	x_{k+1} &= F_{k}x_{k} + w_{k}, \quad 0 \leq k \leq N,
\end{align}
where $k$ denotes the time step, $x_{k} \in \mathbb{R}^{n}$ represents the target motion state, $F_{k} \in \mathbb{R}^{n\times n}$ is the nonsingular state transition matrix.
The process noise $w_{k} \in \mathbb{R}^{n}$ is unknown but bounded. It is assumed to be contained within ellipsoid set $\mathcal{W}_{k}$, satisfying
\begin{align}\label{eq:6}
	\mathcal{W}_{k} = \mathcal{E}_{w_k}(\mathbf{0}, Q_{k}) = \{w_k \in \mathbb{R}^{n} \mid w_{k}^{T}Q_{k}^{-1}w_{k} \leq 1\},
\end{align}
where $Q_{k}$ represents the shape matrix of the ellipsoid set $\mathcal{W}_{k}$, and this matrix is positive definite with appropriate dimensions. 

In this paper, we focus on considering a case of destination constraints, i.e., constraint on the components of the state $x_{N}$ when the target arrives at the destination at time step $N$ and $x_{N}$ satisfies the following linear equation constraint
\begin{align}\label{eq:8}
	Dx_N=d.
\end{align}
Here, the matrix $D\in \mathbb{R}^{m \times n}$ (with $m<n$ and $\text{rank}(D)=m$) and the vector $d \in \mathbb{R}^{m}$ are known. It is important to note that although the destination constraint (\ref{eq:8}) appears to only affect the final state $x_N$, its constraint effect permeates throughout the entire evolution of the target state, since the state updates of the target at each time step are interdependent. 
In fact, using the state equation (\ref{eq:4}), we can establish a recursive relationship between the states $x_k$ and $x_t$ at any two time steps $k$ and $t$, $0\le k\le t\le N$, denoted as
\begin{align}\label{eq:9}
	x_{t}=\Psi_{k,t}x_{k}+\zeta _{k,t},
\end{align}
where
\begin{align}\label{eq:10}
	\Psi_{k,t}=\prod_{i=1}^{t-k}F_{t-i},\quad \Psi_{t,t}=I, \\\label{eq:11}
	\zeta_{k,t}= \begin{cases}
		{\textstyle \sum_{j=k}^{t-1}}\Psi_{j+1,t}w_{j},    & \text{ if } k<t \\
		0, & \text{ others } 
	\end{cases},
\end{align}
and the identity matrix $I$ has appropriate dimensions.
\begin{figure}[htbp]
	\centerline{\includegraphics[width=0.48\textwidth]{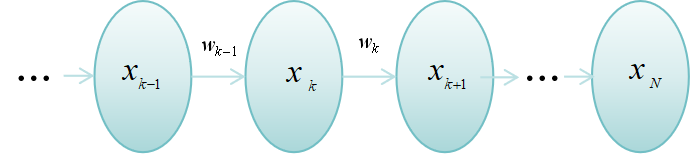}}
	\caption{Unconstrained state evolution.}
	\label{figa1}
\end{figure}
\begin{figure}[htbp]
	\centerline{\includegraphics[width=0.48\textwidth]{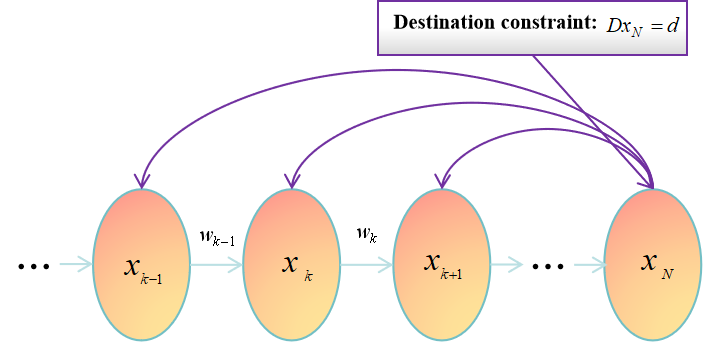}}
	\caption{State evolution with destination constraint.}
	\label{figa2}
\end{figure}
In general, we know that the original state model \eqref{eq:4} is not subject to any constraints acting on it, and when the state transfer matrix is given, the evolution of its state trajectory is only affected by the process noise as shown in Fig. \ref{figa1}.
However, apart from the role  of process noise, if the destination constraint \eqref{eq:8} is imposed on the target state $x_{N}$ at the terminal time step $N$, the entire process of state evolution is also influenced by the role of the destination constraint \eqref{eq:8}, as depicted in Fig. \ref{figa2}.  
Therefore, we now consider how to effectively incorporate the destination constraint information \eqref{eq:8} into the state evolution process to enhance the predictive and control performance of the state model.

\section{Reconstruction of State Models with Destination Constraints}
In the state-space context, state prediction relies on the dynamic model \eqref{eq:4} that depicts the trend of state evolution over time and is used for the generation of state trajectories. Destination constraint information can correct the dynamic model and act as an aid to model prediction. Thus, the state evolution at each time step can be viewed as an exact dynamic model that must obey the destination constraint \eqref{eq:8}. In this section, we consider using convex optimization techniques to project the original state onto the constraints to obtain state decomposition and incorporate the destination constraint information \eqref{eq:8} into the original state model \eqref{eq:4} to complete the reconstruction of the dynamic model.

Specifically, we need to consider how to find an optimal state trajectory with a destination constraint. In other words, we want to find an optimal state trajectory under certain distance or rules while satisfying the destination constraint (\ref{eq:8}) as closely as possible to a state trajectory with a real destination constraint. Noticed that  the original state model \eqref{eq:4} and the destination constraint \eqref{eq:8} serve as two complementary pieces of prior information regarding the evolution of the dynamic system. Consequently, constructing a state model with destination constraint entails skillfully integrating these two pieces of prior knowledge, ensuring that the model is built to facilitate the evolution of the state while automatically satisfying the provided constraint. To maintain the advantages of dealing with constrained linear dynamic systems as outlined in \cite{c30}, we continue to adopt its symbol definitions regarding constrained states and consider the modeling problem of a destination-constrained dynamic system in the set-valued framework. We denote the original constrained combined state as $\mathbf{x}_{k}=[x_{k}^{T}, x_{N}^{T}]^{T} $ and the relaxed constrained combined state as $\mathbf{x}_{k}^{rc}=[(x_{k}^{rc})^{T}, (x_{N}^{rc})^{T}]^{T} $. Let the block matrix $\mathbf{D}=[\mathbf{0}~~ D]\in\mathbb{R}^{m\times 2n}$, where the zero matrix $\mathbf{0}$ has appropriate dimensions. Therefore, the original constraint (\ref{eq:8}) can be rewritten as
\begin{align}\label{eq:12}
	\mathbf{D}\mathbf{x}_{k}=[\mathbf{0}~~ D]\begin{bmatrix}
		x_{k} \\ x_{N}
	\end{bmatrix}=Dx_{N}=d.
\end{align}
For any random vector $z\in\mathbb{R}^{m}$, the corresponding relaxed constraint is represented as
\begin{align}\label{eq:13}
	\mathbf{D}\mathbf{x}_{k}^{rc}=[\mathbf{0}~~ D]\begin{bmatrix}
		x_{k}^{rc} \\ x_{N}^{rc}
	\end{bmatrix}=Dx_{N}^{rc}=z.
\end{align}
Comparing \eqref{eq:12} and \eqref{eq:13}, it can be observed that the states defined here, denoted as $x_{k}$ and $x_{k}^{rc}$, both adhere to the original state model \eqref{eq:4}. The difference lies in the fact that at time step $N$, the corresponding states (designated as $x_{N}$ and $x_{N}^{rc}$, respectively) are subject to different destination constraints. 

Currently, there are many quantified definitions for measuring distance, such as the common $\ell_p$ norms ($p \ge 1$). Using different definitions of distance measurement to calculate the distance between any two points in multi-dimensional space may result in some differences. Thus, choosing an appropriate distance metric is a prerequisite for modeling the state of a constrained system. Mahalanobis distance is a popular choice that has a clear mathematical meaning, which is defined as follow.
For any two vectors $\mathbf{x}\in\mathbb{R}^{n}$ and $\mathbf{y}\in\mathbb{R}^{n}$, their Mahalanobis distance can be defined as $\|\mathbf{x}-\mathbf{y}\|_{W}=\sqrt{(\mathbf{x}-\mathbf{y})^{T}W^{-1}(\mathbf{x}-\mathbf{y})}$, where $W\in\mathbb{R}^{n\times n}$ is a symmetric positive definite matrix. 
Mahalanobis distance is a commonly used distance metric in metric learning, similar to Euclidean distance, utilized for assessing the similarity between data points. It can address issues in high-dimensional linearly distributed data where dimensions are not independently and identically distributed. It can also be seen as a correction to the Euclidean distance, addressing the problem of inconsistent scales and correlations among dimensions. Therefore, this paper also utilizes the Mahalanobis distance to measure the distance between any two states, while simultaneously reflecting the degree of uncertainty among state components.

From a convex optimization perspective, we aim to obtain the optimal state trajectory $\mathbf{x}_{k}^{rc}$ by minimizing the Mahalanobis distance between the relaxed constrained combined state $\mathbf{x}_{k}^{rc}$ and the original constrained combined state $\mathbf{x}_{k}$. This approach naturally incorporates destination constraint information into the state model to reconstruct a state model with destination constraint. In fact, the above problem can be transformed into solving the following optimization problem
\begin{align}\label{eq-15}
	\min_{\mathbf{x}_{k}^{rc}}~~~ &\|\mathbf{x}_{k}^{rc}-\mathbf{x}_{k}\|_{W}^{2}\\\label{eq-16}
	&s.t. ~~ \mathbf{D}\mathbf{x}_{k}^{rc}=d.
\end{align}
The specific description is summarized in the following proposition 1.\\
\textbf{Proposition 1.} 
If we choose the Mahalanobis distance defined as the metric for measuring the distance between any two states, and in \eqref{eq-15} let
\begin{align}\label{eq:30}
	&W=W_{k}=\begin{bmatrix}
		W_{k,1} & W_{k,2} \\
		W_{k,2}^{T} & W_{k,3}
	\end{bmatrix}  \in \mathbb{R}^{2n\times 2n}, W_{k}\succ \mathbf{0}, \\\label{eq:31}
	&\mathbf{D}LW_{N}\mathbf{D}^{T}=\mathbf{0}, L=\begin{bmatrix}
		\mathbf{0} & \mathbf{0}\\
		I & -I
	\end{bmatrix}, 
\end{align}
with $W_{k,\ell}\in \mathbb{R}^{n\times n}, \ell=1,2,3,$ based on the original state model \eqref{eq:4} and the destination constraint \eqref{eq:8}, by solving the optimization problems \eqref{eq-15}-\eqref{eq-16}, then we can derive the decomposition form of the original constrained combined state $\mathbf{x}_{k}$ in \eqref{eq:12} as 
\begin{align}\label{eq:25}
	\mathbf{x}_{k} =A_{k}\mathbf{x}_{k}+(I-A_{k})\mathbf{D}^{\dagger}d,
\end{align}
where $A_{k} = I-W_{k}\mathbf{D}^{T}(\mathbf{D}W_{k}\mathbf{D}^{T})^{-1}\mathbf{D} $, $\mathbf{D}^{\dagger}$ represents the Moore-Penrose pseudoinverse of $\mathbf{D}$, $\mathbf{D}^{\dagger}=\mathbf{D}^{T}(\mathbf{D}\mathbf{D}^{T})^{-1}$.
Furthermore, we can obtain the state model with the destination constraint as 
\begin{align}\label{eq:26}
	x_{k}=\mathbf{F}_{k-1}x_{k-1}+\mathbf{D}_{k-1}+\mathbf{\Xi}_{k-1}w_{k-1}+\sum_{j=k}^{N-1}\mathbf{\Phi}_{j}w_{j},
\end{align}
where
\begin{align*}
	\mathbf{F}_{k-1}&=\mathbf{\Xi}_{k-1}F_{k-1},\\
	\mathbf{\Xi}_{k-1}&=I-\mathbf{B}_{k-1}\Psi_{k,N}, \\
	\mathbf{B}_{k-1}&=W_{k,2}D^{T}(DW_{k,3}D^{T})^{-1}D, \\
	\mathbf{D}_{k-1}&=\mathbf{B}_{k-1}D^{\dagger}d, \\
	D^{\dagger}&=D^{T}(DD^{T})^{-1}, \\
	\mathbf{\Phi}_{j}&=-\mathbf{B}_{k-1}\Psi_{j+1,N}, j=k,\dots,N-1.
\end{align*}

\textbf{Proof:} See Appendix A.

\textit{Remark 1.
(i) Actually, the $A_{k}$ in \eqref{eq:25} is a projection operator onto the null space $\mathcal{N}(\mathbf{D})$ satisfying $\mathbf{D}A_{k}=\mathbf{0}$. Left-multiplying both sides of equation \eqref{eq:25} by the matrix $\mathbf{D}$ yields $\mathbf{D}\mathbf{x}_{k}=\mathbf{D}A_{k}+\mathbf{D}(I-A_{k})\mathbf{D}^{\dagger}d=d$, which shows that the decomposition of the combined state $\mathbf{x}_{k}$ obeys the original constraint \eqref{eq:12}. (ii) Let $k=N$ in \eqref{eq:26}, and left-multiply both sides of \eqref{eq:26} by a matrix $D$, in conjunction with the conditions of Proposition 1, $\mathbf{D}LW_{N}\mathbf{D}^{T}=D(W_{N,2}-W_{N,3} )D^{T}=\mathbf{0}$. It is easy to verify that the reconstructed state model \eqref{eq:26} adheres to the original destination constraint \eqref{eq:8}. This indicates that we have successfully incorporated the destination constraint information \eqref{eq:8} into the state model. (iii) The model \eqref{eq:26} constructed for the state with destination constraint does not utilize information about the process noise distribution, and therefore, it holds for any noise distribution without being constrained by assumptions about the noise distribution, such as Gaussian distribution, uniform distribution, and so forth. Moreover, the computational process for decomposing the original constraint combination state \eqref{eq:25} only necessitates the linearity of the constraint; it does not have a specific dependency on the type of state model, whether it is linear or nonlinear. }

It is worth noting that when the original state model \eqref{eq:4} and the destination constraint information \eqref{eq:8} are given, then the values of all model parameters ($\mathbf{F}_{k-1}$, $\mathbf{D}_{k-1}$, $\mathbf{\Xi}_{k-1}$ and $\mathbf{\Phi}_{j}, j=k,\cdots,N-1$) of the reconstructed state model \eqref{eq:26} with destination constraint are determined by the weight matrix $W_{k}$. The evolution process of the state trajectory varies based on the selection of $W_{k}$. Effective prediction and control of the state trajectory can only be obtained through the utilization of the weight matrix. One strategy involves the minimization of process noise generated at each time step, aiming to reduce it to the most feasible extent within the domain of state control. This endeavor serves to enhance the accuracy of estimating the target's motion state, promoting a more precise tracking of the target's position. Minimizing process noise also improves the system's robustness, lowering sensitivity to model mismatches and external interferences, thereby ensuring reliable target tracking in uncertain environments. 

\section{Selection of the Weight Matrix}
Different choices of the weight matrix $W_{k}$ in \eqref{eq:30} may lead to non-uniform effects of destination constraint \eqref{eq:8} on the state evolution at each time step. This, in turn, may result in irregular jumps or multi-peak in the trend of the motion trajectory generated by the state model \eqref{eq:26}, leading to a lack of smoothness in the trajectory, which does not align with the actual law of the target's motion. Hence, the main objective of this section is to establish the optimization problem for the selection of the weight matrix $W_{k}$ as stated in Proposition 1. This is obtained by minimizing the process noise in the context of set-valued scenarios. By bounding the combined vector of process noise, we carefully design the optimal weight matrix $W_{k}^{*}$ for this optimization problem. The optimality theory for $W_{k}^{*}$ is formulated in the form of a proposition.
\subsection{Problem Formulation}
For the sake of descriptive convenience, we let  \begin{align}\nonumber
	\mathbf{\Psi}_{k}&=[\Psi_{k,N},\cdots,\Psi_{N,N}],  \mathbf{w}_{k-1}=(w_{k-1}^{T},\cdots,w_{N-1}^{T})^{T},\\\label{eq16}
	\mathbf{w}_{0}&=(w_{0}^{T},\cdots,w_{N-1}^{T})^{T},
	\mathbf{G}_{k}=[I,\underbrace{\mathbf{0},\cdots,\mathbf{0}}_{\text{A total of}~ N-k}],\\\nonumber
	\mathbf{C}_{k}&=[\mathbf{G}_{k}^{T}, \mathbf{\Psi}_{k}^{T}]^{T},~\text{and}~
	H_{k}=\mathbf{G}_{k}-\mathbf{B}_{k-1}\mathbf{\Psi}_{k}.
\end{align}
The definitions of $\Psi_{i,N}$ for $i=k,\cdots,N$, $w_{j}$ for $j=0,\cdots, N-1$, and $\mathbf{B}_{k-1}$ can be viewed in \eqref{eq:10}, \eqref{eq:6} and \eqref{eq:26} respectively. According to \eqref{eq16}, $\mathbf{w}_{k-1}$ can be obtained from 
$\mathbf{w}_{0}$
through the following affine transformation 
\begin{align}\label{eq59}
	\mathbf{w}_{k-1}=M_{k}\mathbf{w}_{0}.
\end{align}
The matrix 
$M_{k}\in\mathbb{R}^{(N-k)\times N}$
satisfies that the element in the $i$th row and 
$(k-1+i)$th column is 1, $i=1,\cdots,N-k$, while all other elements are 0, that is, 
\begin{align}\label{eq60}
	M_{k}=\begin{bmatrix}
		0 & \cdots & 0 & 1 & 0 & 0 & \cdots & 0\\
		0 & \cdots & 0 & 0 & 1 & 0 & \cdots & 0\\
		0 & \cdots & 0 & 0 & 0 & 1 & \cdots & 0\\
		\vdots & \cdots & \vdots & \vdots & \vdots & \vdots & \cdots & \vdots\\
		0 & \cdots & 0 & 0 & 0 & 0 & \cdots & 1
	\end{bmatrix}_{{(N-k)\times N}}.
\end{align}

Given the consideration of process noises in this paper in the unknown but bounded setting, a more specific assumption is made that the combined noise vector $\mathbf{w}_{0}$ belongs to a known bounded ellipsoidal set, that is,
\begin{align}\label{eq106}
\mathbf{w}_{0}\in\mathcal{W}_{\mathbf{w}_{0}},
\end{align}
where 
\begin{align}\label{eq63}
	\mathcal{W}_{\mathbf{w}_{0}}:=\mathcal{W}_{\mathbf{w}_{0}}(\mathbf{0},Q_{\mathbf{w}_{0}})=\{\mathbf{w}_{0}\in\mathbb{R}^{N}|\mathbf{w}_{0}^{T}Q_{\mathbf{w}_{0}}^{-1}\mathbf{w}_{0}\le 1\}
\end{align}
with the zero vector $\mathbf{0}$ as the center and a positive definite shape matrix 
\begin{align}\label{eq23}
	Q_{\mathbf{w}_{0}}=\begin{bmatrix}
		Q_{0} & Q_{0,1} & \cdots & Q_{0,N-1}\\
		Q_{1,0} & Q_{1} & \cdots & Q_{1,N-1}\\
		\vdots & \vdots & \ddots & \vdots\\
		Q_{N-1,0} & Q_{N-1,1} & \cdots & Q_{N-1}
	\end{bmatrix},
\end{align}
in which $Q_{i}=Q_{i,i}\in \mathbb{R}^{n\times n} $ and $Q_{i,j} \in \mathbb{R}^{n\times n}, i\ne j, i,j=0,\cdots,N-1$. 

Considering the context of set-valued scenarios, based on the reconstruction model \eqref{eq:26} and condition \eqref{eq106}, calculate the coverage ellipsoid of process noise introduced when updating the state from $x_{k-1}$ to $x_{k}$. According to \eqref{eq:26}, \eqref{eq16}, and \eqref{eq59}, we can express the term associated with the noise as 
\begin{align}\nonumber
	\eta_{k} & := \mathbf{\Xi}_{k-1}w_{k-1}+\sum_{j = k}^{N-1}\mathbf{\Phi}_{j}w_{j}\\\nonumber
	& = (\mathbf{G}_{k}-\mathbf{B}_{k-1}\mathbf{\Psi}_{k})\mathbf{w}_{k-1}\\\label{eq74}	
	&=H_{k}M_{k}\mathbf{w}_{0},
\end{align} 
By \eqref{eq106} and \eqref{eq74}, using the affine invariance of the ellipsoid, one can obtain an affine ellipsoid $\mathcal{W}_{\eta_{k}}:=\mathcal{W}_{\eta_{k}}(\mathbf{c}_{\eta_{k}},Q_{\eta_{k}})=H_{k}M_{k}\mathcal{W}_{\mathbf{w}_{0}}$ of $\mathcal{W}_{\mathbf{w}_{0}}$ to cover the variable $\eta_{k}$, and the center and shape matrices corresponding to the two ellipsoids satisfy the following relations:  \begin{align}\label{eq76}
	\mathbf{c}_{\eta_{k}}&=H_{k}M_{k}\cdot\mathbf{0}=\mathbf{0},\\\label{eq84}
	Q_{\eta_{k}}&=H_{k}M_{k}Q_{\mathbf{w}_{0}}M_{k}^{T}H_{k}^{T}=H_{k}Q_{\mathbf{w}_{k-1}}H_{k}^{T},
\end{align}
with 
\begin{align}\label{eq85}
	Q_{\mathbf{w}_{k-1}}=M_{k}Q_{\mathbf{w}_{0}}M_{k}^{T}.
\end{align}

To mitigate the uncertainties arising from the utilization of model \eqref{eq:26} with arbitrary weight matrice $W_{k}=\begin{bmatrix}
	W _{k,1} & W_{k,2} \\
	W_{k,2}^{T} & W_{k,3}
\end{bmatrix}\succ \mathbf{0}$ and during state evolution, we consider minimizing the shape matrix of the ellipsoid covering the process noise generated at each time step. This endeavor seeks to derive the optimal weight matrix $W_{k}^{*}$. It can be formulated as the optimization problem:
\begin{align}\label{eq73}
	\min_{W_{k}}& ~~ Q_{\eta_{k}}(W_{k})\\\nonumber
	&\text{s.t.}~~ 
	W_{k} \succ \mathbf{0}, 
\end{align}
where the shape matrix
\begin{align*}
&Q_{\eta_{k}}(W_{k}):=Q_{\eta_{k}}\\
&=(\mathbf{G}_{k}-\mathbf{I}_{1}W_{k}\mathbf{I}_{3}D^{T}(D\mathbf{I}_{2}W_{k}
\mathbf{I}_{3}D^{T})^{-1}D\mathbf{\Psi}_{k})M_{k}Q_{\mathbf{w}_{0}}M_{k}^{T}\\
&~~~~(\mathbf{G}_{k}-\mathbf{I}_{1}W_{k}\mathbf{I}_{3}D^{T}(D\mathbf{I}_{2}W_{k}
\mathbf{I}_{3}D^{T})^{-1}D\mathbf{\Psi}_{k})^{T}
\end{align*}
 with $\mathbf{I}_{1}=\begin{bmatrix}
	I & \mathbf{0}
\end{bmatrix}$, $\mathbf{I}_{2}=\begin{bmatrix}
	\mathbf{0} & I
\end{bmatrix}$, and $\mathbf{I}_{3}=\begin{bmatrix}
	\mathbf{0} \\
	I
\end{bmatrix}$. The derivation is due to \eqref{eq84}, \eqref{eq16} and \eqref{eq:26}. Here $\mathbf{G}_{k}$ and $\mathbf{\Psi}_{k}$ are defined in \eqref{eq16}, and the definitions of $M_{k}$ and  $Q_{\mathbf{w}_{0}}$ can be seen in \eqref{eq60} and \eqref{eq23}, respectively. 

It can be observed that \eqref{eq73} represents a matrix optimization problem. We provide an optimal solution and a proof of its optimality in next subsection.
\subsection{Optimal Weight Matrix Design}

By minimizing the shape matrices of the corresponding ellipsoids covering the process noise at each time step, we can derive the analytic expression for the optimal weight matrix $W_{k}^{*}$ in \eqref{eq73}, and this conclusion is specifically summarized in the following Proposition 2. \\
\textbf{Proposition 2.} If the noise vector $\mathbf{w}_{0}$ in a bounded ellipsoid, i.e., \eqref{eq106} holds, and the state transfer matrix $F_{k-1}$ is given, then an optimal solution of \eqref{eq73} is given by:
\begin{align}\label{eq111}
	W_{k}^{*}&=\begin{bmatrix}
		Q_{k-1} & \mathbf{G}_{k}Q_{\mathbf{w}_{k-1}}\mathbf{\Psi}_{k}^{T} \\
		\mathbf{\Psi}_{k}Q_{\mathbf{w}_{k-1}}\mathbf{G}_{k}^{T} & \mathbf{\Psi}_{k}Q_{\mathbf{w}_{k-1}}\mathbf{\Psi}_{k}^{T}
	\end{bmatrix},\\\label{eq112} 
	W_{k,1}^{*}&=Q_{k-1},\\\label{eq109}
	W_{k,2}^{*}&=\sum_{i=k-1}^{N-1}Q_{k-1,i}\Psi_{i+1,N}^{T},\\\label{eq113} 
	W_{k,3}^{*}&=\sum_{j=k-1}^{N-1}
	\sum_{i=k-1}^{N-1}\Psi_{i+1,N}Q_{i,j}\Psi_{j+1,N}^{T},
\end{align}
where $\mathbf{G}_{k}$ and $\mathbf{\Psi}_{k}$ are defined in \eqref{eq16}. The definitions of $Q_{\mathbf{w}_{k-1}}$ and $\Psi_{i+1,N}, i=k-1,\cdots,N-1$ can be viewed in \eqref{eq85} and \eqref{eq:10}, respectively.
Furthermore, we can infer that the weight matrix $W_{k}^{*}=\begin{bmatrix}
	W_{k,1}^{**} & W_{k,2}^{*}\\
	(W_{k,2}^{*})^{T} & W_{k,3}^{*}
\end{bmatrix}\succ \mathbf{0}$ is also an optimal solution to \eqref{eq73}, where $W_{k,1}^{**}\succ \mathbf{0}$, $W_{k,1}^{**}-W_{k,2}^{*}(W_{k,3}^{*})^{-1}(W_{k,2}^{*})^{T}\succ \mathbf{0}$, and the expressions for $W_{k,2}^{*}$ and $W_{k,3}^{*}$ are given by \eqref{eq109} and \eqref{eq113}, respectively. 

\textbf{Proof:} See Appendix B.

\textit{Remark 2. (i) The conclusion of Proposition 2 suggests that when transitioning from the state at time step $k-1$ to the state at time step $k$, utilizing the reconstructed state model \eqref{eq:26}, and employing the weight matrix $W_{k}^{*}$ in \eqref{eq111}, results in the minimal process noise. This provides optimality theory guarantees for the selection of weight matrix $W_{k}$ for model \eqref{eq:26} in the set-valued framework. It should be noted that the optimal solution $W_{k}^{*}$ for the optimization problem \eqref{eq73} is not unique. This is due  to the value of $W_{k,1}^{**}$ is undetermined, and its value only needs to ensure that $W_{k}^{*}$ is positive definite. In this case, for the model \eqref{eq:26}, selecting the sub-block matrices $W_{k,2}^{*}$ and $W_{k,3}^{*}$ as designed in Proposition 2 results in obtaining the optimal weight matrix $W_{k}^{*}$. (ii) From Proposition 2, we deduced that $Q_{\eta_{k}}(W_{k}^{*})\preceq Q_{\eta_{k}}(W_{k})$, where arbitrary weight matrice $W_{k} \succ \mathbf{0}$. Also, we know that the size of the ellipsoid can be expressed as a metric function of the shape matrix, denoted as $\Omega(\cdot)$, as detailed in references \cite{cc1} and \cite{cc2}. If we choose commonly used trace or $\log\det$ functions as the metric $\Omega(\cdot)$ and leverage basic algebraic knowledge, we can derive a scalar-form result, namely $\Omega(Q_{\eta_{k}}(W_{k}^{*}))\preceq \Omega(Q_{\eta_{k}}(W_{k}))$, with $\Omega(\cdot)=\text{tr}(\cdot)$ or $\Omega(\cdot)=\log\det(\cdot)$.}

Proposition 2 utilizes the known bounded ellipsoid 
$\mathcal{W}_{\mathbf{w}_{0}}$ to cover the composite noise vector $\mathbf{w}_{0}$. Based on the convex optimization problem \eqref{eq73}, an expression for the weight matrix satisfying the constraints is given and proved to be optimal in the set-valued context.

\section{Theoretical properties of reconstruction model}
In this section, we explore the theoretical properties of the state model \eqref{eq:26} when equipped with optimal weight matrix.

In fact, the constructed state model \eqref{eq:26} yields smaller process noise at time $k$ when transitioning from state $x_{k-1}$ to $x_k$ compared to the original state model \eqref{eq:4}, i.e., the unconstrained state model. The specific results are provided by Proposition 3.\\
\textbf{Proposition 3.} State trajectories are generated using the state model \eqref{eq:26} with optimal weight matrix $W_{k}^{*}$ in \eqref{eq111} and initial state model \eqref{eq:4}, respectively. If the conditions of Proposition 2 hold, then the size of the bounded covering ellipsoid corresponding to the process noise generated using model \eqref{eq:26} when updating the state from $x_{k-1}$ to $x_{k}$ is smaller than the one generated by model \eqref{eq:4}, i.e., 
\begin{align}\label{eq91}
	Q_{\eta_{k}}(W_{k}^{*})\preceq  Q_{k-1},
\end{align}
where $Q_{\eta_{k}}(W_{k}^{*})=H_{k}^{*}Q_{\mathbf{w}_{k-1}}(H_{k}^{*})^{T}$ and $Q_{k-1}$ denote the shape matrices of the ellipsoids corresponding to the process noise generated by models \eqref{eq:26} and \eqref{eq:4}, respectively. Equation \eqref{eq85} gives the definition of $Q_{\mathbf{w}_{k-1}}$. The matrices $H_{k}^{*}=\mathbf{G}_{k}-\mathbf{B}_{k-1}^{*}\mathbf{\Psi}_{k}$ and $\mathbf{B}_{k-1}^{*}=W_{k,2}^{*}D^{T}(DW_{k,3}^{*}D^{T})^{-1}D$, where $\mathbf{G}_{k}$ and $\mathbf{\Psi}_{k}$ are defined in \eqref{eq16}. 

\textbf{Proof:} See Appendix C.

Proposition 3 shows that in the set-valued framework, using the state model \eqref{eq:26} with the optimal weight matrix results in process noise at time step $k$ that is necessarily smaller than that generated by the original unconstrained state model \eqref{eq:4}. This indicates that our reconstructed state model \eqref{eq:26} effectively incorporates destination constraint information, leading to reduced system bias and enhanced control  and predictive performance of the state model.

As analyzed in Section II, the presence of dependence in the state evolution process implies that the destination constraint influences the state at each time step, denoted as $x_{k}$. In particular, if the matrix $D$ in \eqref{eq:8} is invertible, which means it imposes constraints on all components of the terminal state $x_{N}$ at time $N$, then we can state the following proposition.\\
\textbf{Proposition 4.} If the model parameters of the initial state model \eqref{eq:4} are time-invariant and the shape matrix $Q_{\mathbf{w}_{0}}$ of the ellipsoid covering the vector $\mathbf{w}_{0}$ is a time-invariant diagonal matrix, i.e., $Q_{\mathbf{w}_{0}}=\text{diag}(Q,Q,\cdots,Q)$, then we have 
\begin{align}\label{eq97}
	Q_{\eta_{k+1}}(W_{k+1}^{*})\preceq  Q_{\eta_{k}}(W_{k}^{*}).
\end{align}
Here $Q_{\eta_{k}}(W_{k}^{*})=H_{k}^{*}Q_{\mathbf{w}_{k-1}}(H_{k}^{*})^{T}$ and $Q_{\eta_{k+1}}(W_{k+1}^{*})=H_{k+1}^{*}Q_{\mathbf{w}_{k}}(H_{k+1}^{*})^{T}$ represent the shape matrices corresponding to the covering ellipsoids of the process noise generated at time steps $k$ and $k+1$ using the state model \eqref{eq:26}, respectively. The definition of $Q_{\mathbf{w}_{k-1}}$ can be found in \eqref{eq85}. In this context, $\mathbf{B}_{k}^{*}=W_{k+1,2}^{*}D^{T}(DW_{k+1,3}^{*}D^{T})^{-1}D$, and $H_{k+1}^{*}=\mathbf{G}_{k+1}-\mathbf{B}_{k}^{*}\mathbf{\Psi}_{k+1}$ with $\mathbf{G}_{k+1}$ and $\mathbf{\Psi}_{k+1}$ are defined in \eqref{eq16}. 

\textbf{Proof:} See Appendix D.

Proposition 4 indicates that when generating state trajectories using the refactored state model \eqref{eq:26}, the constraints acting on the target motion state become increasingly significant over time as it gets closer to the destination. Consequently, this results in a reduction of uncertainty introduced, implying that the impact of process noise on the state evolution diminishes as the target approaches its destination.

\section{Simulation Experiments and Discussions}
Consider a flying target in a 2D scenario that starts at coordinates (0 km, 10 km) with an initial velocity of (240 m/s, 0 m/s). After flying for 50 seconds, it reaches its destination at coordinates (12 km, 0 km). At the time step of arrival, the angle between the target's heading and the horizontal direction is denoted as $\theta$, measured in a clockwise direction from due east.
According to the construction of $W_{k}^{*}$ in \eqref{eq111}, we aim to obtain a more realistic state evolution pattern by controlling the state transition matrix and incorporating process noise information. This results in state updates that are influenced by the destination constraint in a relatively uniform manner, while also exhibiting randomness. For computational convenience, we pick $Q_{\mathbf{w}_{0}}=\text{diag}(Q_{0},\cdots,Q_{N-1}) $. By Proposition 2, the expressions for $W_{k,2}$ and $W_{k,3}$ in model \eqref{eq:26} are as follows: 
\begin{align*}
	W_{k,2}=Q_{k}\Psi_{k,N}^{T}, W_{k,3}=\sum_{j=k}^{N}\Psi_{j,N}Q_{k}\Psi_{j,N}^{T},
\end{align*} 
respectively.
Furthermore, we assume that the relaxed state model is an almost constant velocity (CV) dynamic model: 
\begin{align}\label{eq44}
	x_{k+1}^{rc}=F_{k}x_{k}^{rc}+w_{k},
	w_{k}\in\mathcal{W}_{k}=\mathcal{E}_{w_{k}}(\mathbf{0},Q_{k}).
\end{align}
The state vector $x^{rc}=(x,y,\dot{x},\dot{y}) $ is composed of position components $(x,y) $ and velocity components $(\dot{x},\dot{y}) $, with
\begin{align*}
	F_{k}=\begin{bmatrix}
		1 & T & 0 & 0\\
		0 & 1 & 0 & 0\\
		0 & 0 & 1 & T\\
		0 & 0 & 0 & 1
	\end{bmatrix},
	Q_{k}=g^{2}\begin{bmatrix}
		\frac{T^{3}}{3} & \frac{T^{2}}{2} & 0 & 0\\
		\frac{T^{2}}{2} & T & 0 & 0\\
		0 & 0 & \frac{T^{3}}{3} & \frac{T^{2}}{2}\\
		0 & 0 & \frac{T^{2}}{2} & T
	\end{bmatrix},
\end{align*}
where $T=1$ s represents the sampling time interval, and $g=9.8 $ m/s² is the gravitational acceleration. The parameters in destination constraint \eqref{eq:8} are set with the following values: total sampling time $N=50 $, 
\begin{align*}
	D=\begin{bmatrix}
		1 & 0 & 0 & 0\\
		0 & 1 & 0 & 0\\
		0 & 0 & 1 & -\text{cot}(\theta)
	\end{bmatrix}~\text{and}~
	d=\begin{bmatrix}
		12000 \\
		0 \\
		0
	\end{bmatrix}.
\end{align*}

\begin{figure}[htbp]
	\centerline{\includegraphics[width=0.5\textwidth]{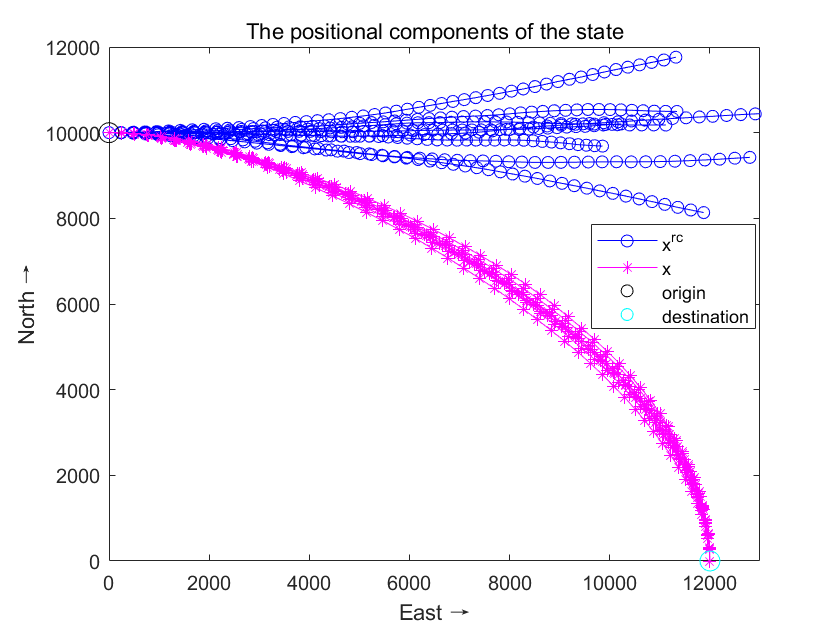}}
	\caption{The trajectories of $\mathbf{x}^{rc} $ and $\mathbf{x} $, with $\theta=90^{\circ}$ at the destination.}
	\label{fig2}
\end{figure}

\begin{figure}[htbp]
	\centerline{\includegraphics[width=0.5\textwidth]{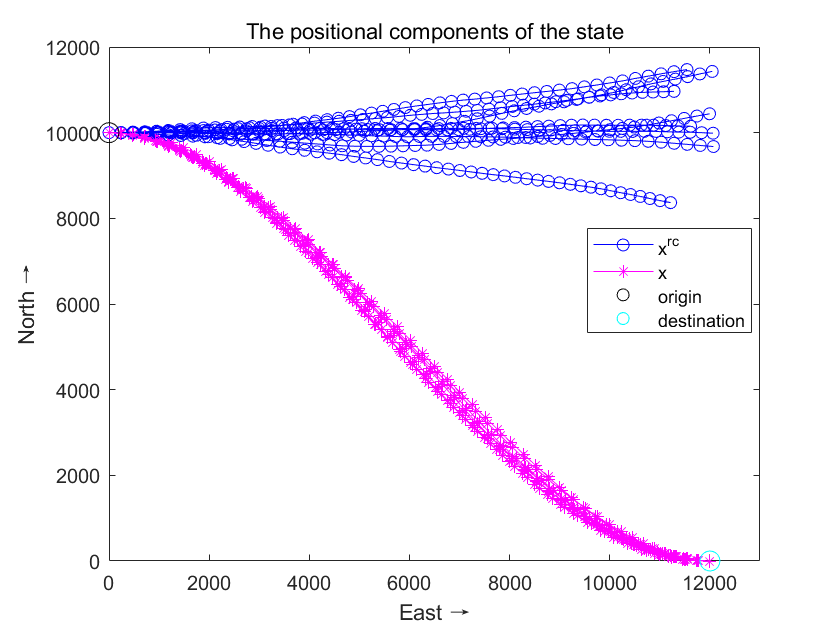}}
	\caption{The trajectories of $\mathbf{x}^{rc} $ and $\mathbf{x} $, with $\theta=0^{\circ}$ at the destination.}
	\label{fig3}
\end{figure}

Substituting these parameter settings into the state model \eqref{eq:26} with destination constraint and the relaxed (almost CV) state model \eqref{eq44}, we randomly generated eight trajectories of state evolution for each model. These trajectories are depicted in Fig. \ref{fig2} ($\theta=90^{\circ}$ at the destination) and Fig. \ref{fig3} ($\theta=0^{\circ}$ at the destination). In both figures, two state models have generated two sets of state trajectories. Although they have the same origin, their trajectory evolution shows significant differences. Notably, trajectories generated by the relaxed state model, denoted as $\mathbf{x}^{rc}$, exhibit a scattered and random evolution trend, and none of these trajectories pass through the known destination. In contrast, the destination-constrained state model generates trajectories denoted as $\mathbf{x} $, displaying a more consistent trend of progressing towards the destination. Importantly, all of these trajectories are guaranteed to pass through the known destination. This validates that our constructed destination-constrained state model incorporates destination constraint information into the state evolution process. Moreover, all generated trajectories are smooth and natural, indicating that the weight matrix $W_{k}$ we designed is appropriate. It effectively applies the destination constraint uniformly to each time step of state evolution, enhancing the predictive capability of the state model.  

\begin{figure}[htbp]
	\centerline{\includegraphics[width=0.5\textwidth]{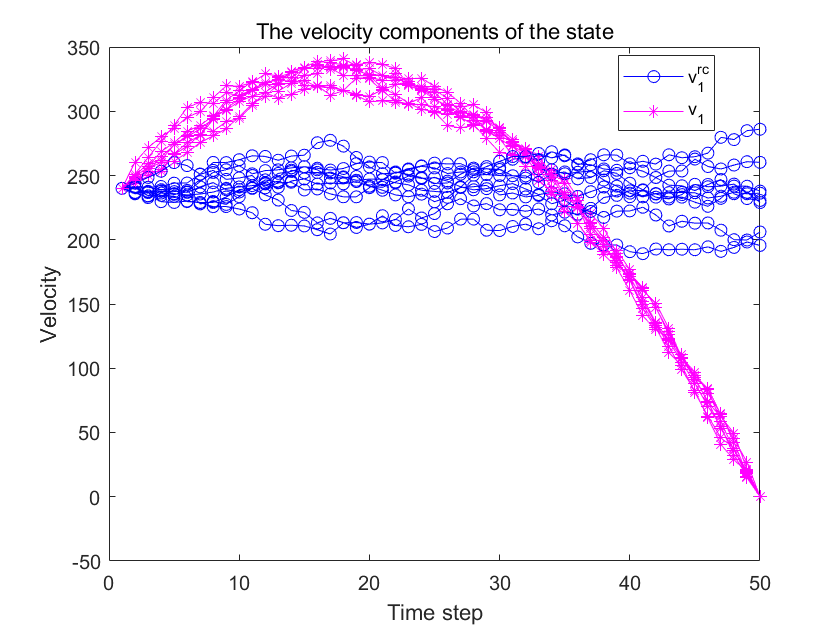}}
	\caption{The curves of  velocity components (eastward) of $x_{k}^{rc}$ and $x_{k} $, $\theta=90^{\circ}$ at the destination.}
	\label{fig4}
\end{figure}

\begin{figure}[htbp]
	\centerline{\includegraphics[width=0.5\textwidth]{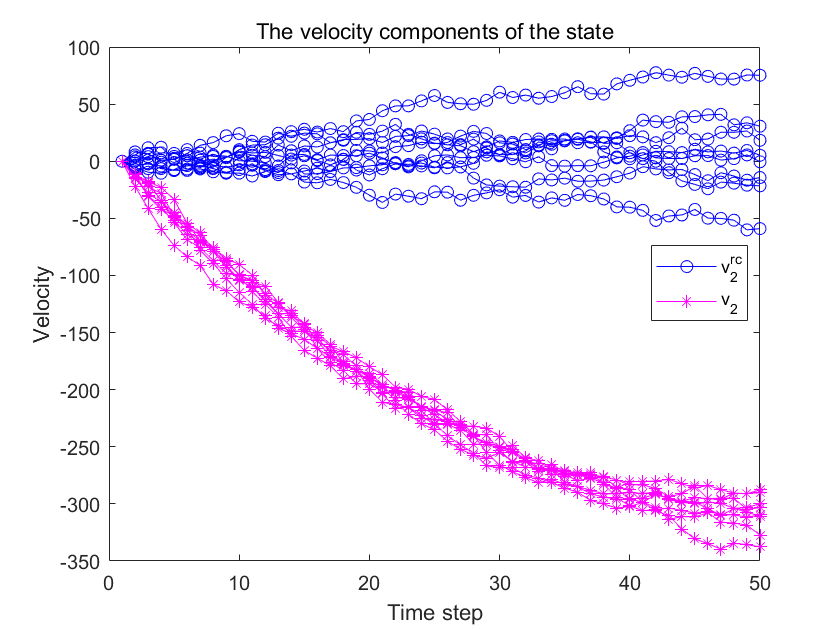}}
	\caption{The curves of  velocity components (northward) of $x_{k}^{rc}$ and $x_{k} $, $\theta=90^{\circ}$ at the destination.}
	\label{fig5}
\end{figure}

\begin{figure}[htbp]
	\centerline{\includegraphics[width=0.5\textwidth]{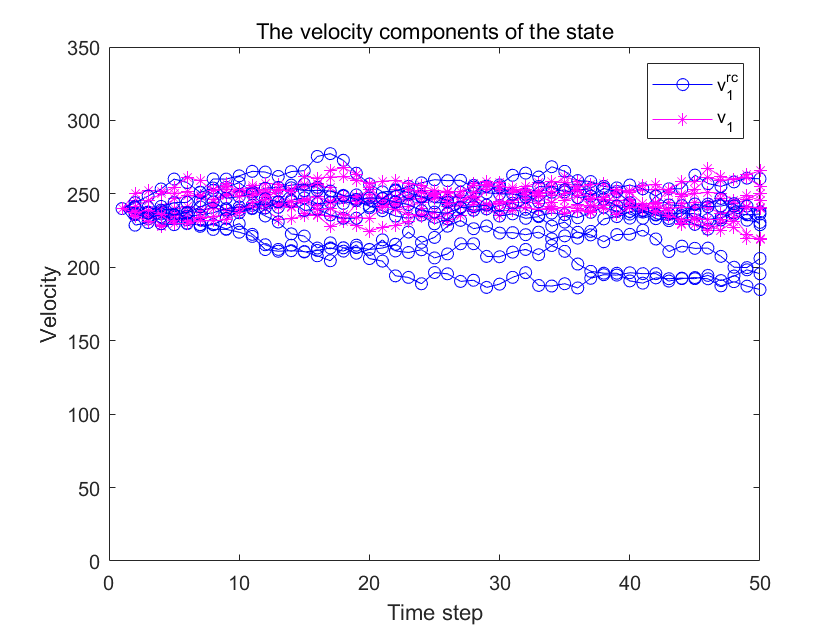}}
	\caption{The curves of  velocity components (eastward) of $x_{k}^{rc}$ and $x_{k} $, $\theta=0^{\circ}$ at the destination.}
	\label{fig6}
\end{figure}

\begin{figure}[htbp]
	\centerline{\includegraphics[width=0.5\textwidth]{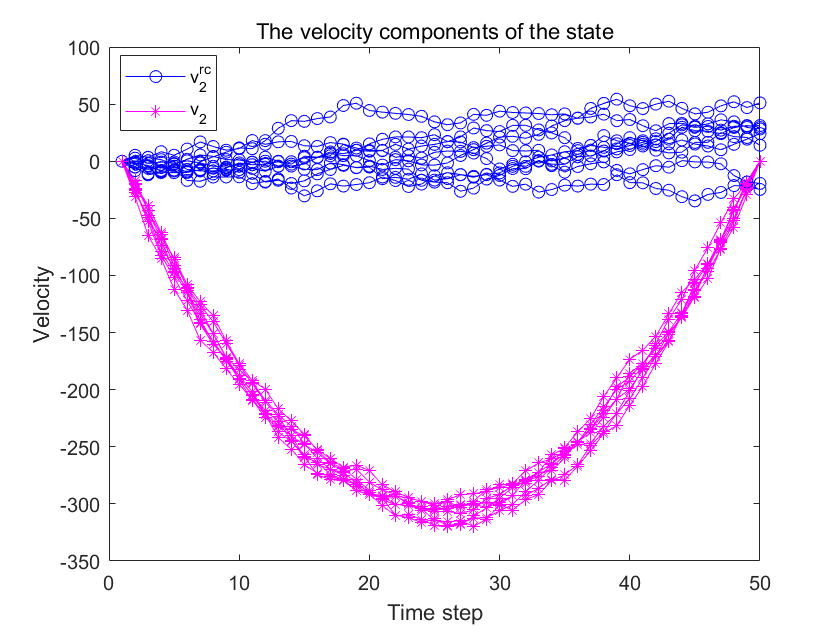}}
	\caption{The curves of  velocity components (northward) of $x_{k}^{rc}$ and $x_{k} $, $\theta=0^{\circ}$ at the destination.}
	\label{fig7}
\end{figure}

When reaching the destination with a heading angle $\theta=90^{\circ}$ relative to the horizontal direction, we have plotted two sets (each set containing 8 curves) of eastward velocity component $\text{v}_{1}^{rc} $ and $\text{v}_{1} $ corresponding to states $x_{k}^{rc} $ and $x_{k} $, as shown in Fig. \ref{fig4} and Fig. \ref{fig5}. Similarly, when $\theta=0^{\circ}$, we have also plotted the curves of eastward and northward velocity components for states $x_{k}^{rc} $ and $x_{k}$, as shown in Fig. \ref{fig6} and Fig. \ref{fig7}. Fig. \ref{fig4}-\ref{fig7} reflects the variation in velocity in different directions for various models when different values of $\theta$ are fixed.


\begin{figure}[htbp]
	\centerline{\includegraphics[width=0.5\textwidth]{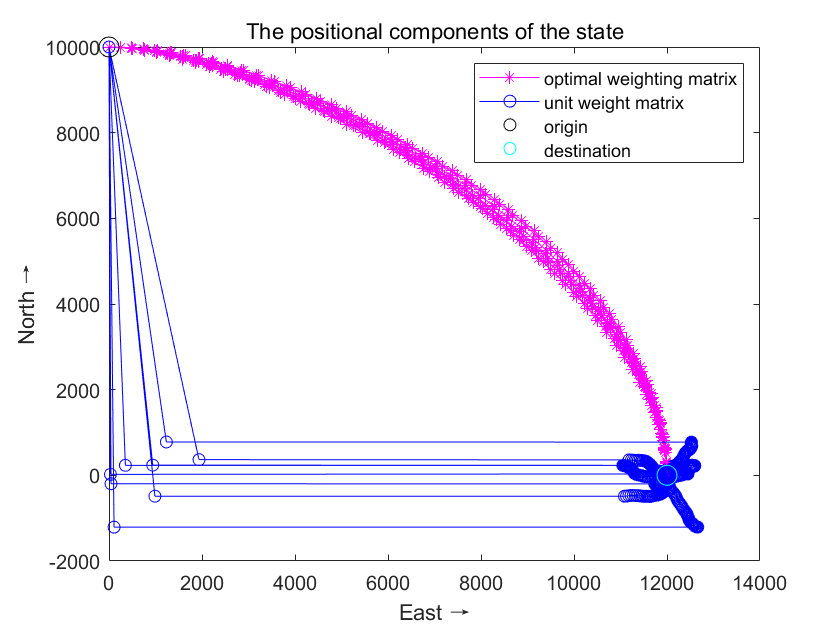}}
	\caption{Compare the trajectory evolution of $\mathbf{x}$ with the optimal weight matrix and the identity weight matrix, with $\theta=90^{\circ}$ at the destination.}
	\label{fig8}
\end{figure}

For \eqref{eq:26}, we chose two different scenarios: one with $W_{k,2}=W_{k,3}=I$ (unoptimized weights), where $I$ is the appropriately sized identity matrix, and another using the optimal weight matrix designed in Proposition 2. We plotted the evolution trajectories of states $x_{k}$ with destination constraints in both cases, generating 8 random trajectories for each scenario. The results are depicted in Fig. \ref{fig8}. In Fig. \ref{fig8}, it is evident that trajectories generated using the identity weight matrix exhibit pronounced jumping behavior and follow irregular trends in orienting towards the destination. Such trajectories are inadequate for practical applications. On the contrary, trajectories produced with the optimal weight matrix consistently maintain a trend of orienting towards the destination. Moreover, these trajectories appear smooth and natural. This indicates that the model with the optimized weight matrix has a clear advantage over the model with the unoptimized weight matrix.

\begin{figure}[htbp]
	\centerline{\includegraphics[width=0.5\textwidth]{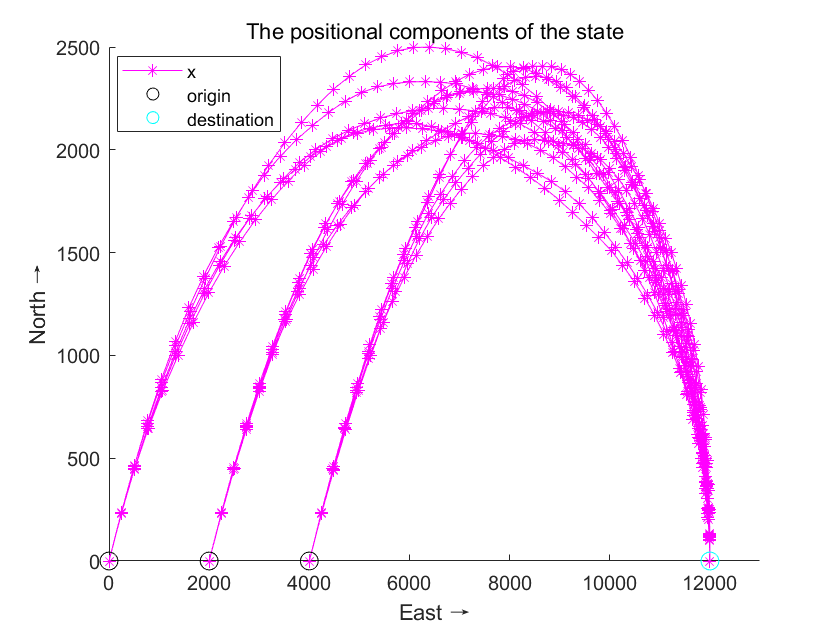}}
	\caption{Three sets of $\mathbf{x}$ trajectories generated by three different origins but converging to the same destination, and $\theta=90^{\circ}$ at the destination.}
	\label{fig9}
\end{figure}

\begin{figure}[htbp]
	\centerline{\includegraphics[width=0.5\textwidth]{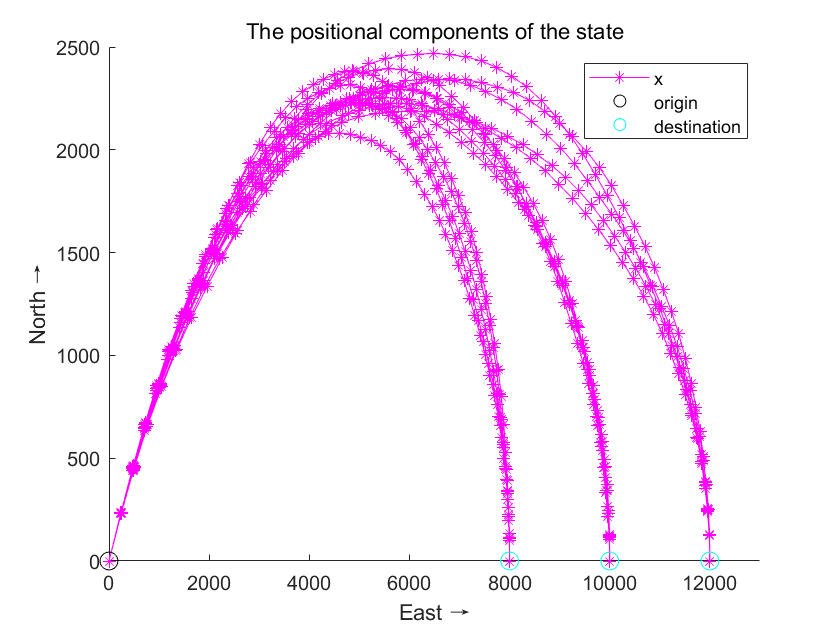}}
	\caption{Three sets of $\mathbf{x}$ trajectories generated from the same origin but constrained by three different destinations, and $\theta=90^{\circ}$ at the destination.}
	\label{fig10}
\end{figure}

Fig. \ref{fig9} illustrates that when $\theta=90^{\circ}$, three distinct sets of trajectory patterns emerge from different starting points. Each set consists of 5 randomly generated trajectories constrained by the common destination. Despite the variability in starting points, all trajectories converge to the known destination due to the shared destination constraint. In Fig. \ref{fig10}, with $\theta=90^{\circ}$, the destination originates from the same initial point. However, three sets of distinct trajectory patterns are generated, each comprising 5 randomly generated trajectories. This diversity arises from the constraints imposed by three different positions of the destination. As a result, all trajectories, guided by their specific destination constraints, eventually pass through their respective constrained destinations.

\section{Conclusion}
In the stochastic framework, modeling dynamic systems with constrained information has long been a subject of extensive research and has yielded rich results. However, within the realm of the set-valued framework, especially when dealing with dynamic systems that incorporate destination constraints, the research has been relatively limited. This paper delves into the reconstruction and associated theoretical properties of linear dynamic system model with destination constraint in the set-valued framework, yielding valuable outcomes.

We introduce a novel approach by quantifying the destination constraint as an integral element of preexisting information that governs state evolution. This involves seamlessly integrating the constraint into the entire sequence of state transitions. This innovative technique, rooted in the set-valued framework, utilizes convex optimization projections to enhance its effectiveness. The result is a coherent and comprehensive destination-constrained model that adeptly accounts for the nuanced influence of destination constraints on state evolution in real-time. To further refine our model's performance, we introduce an weight matrix that facilitates smooth and natural state trajectory progression. This dynamic matrix ensures that state trajectories unfold organically, aligning their evolutionary patterns more closely with reality. Continuing in the set-valued framework, we provide an analytical expression for the weight matrix of the reconstructed model, establish its optimality theory, and explore the associated theoretical properties of the model. Our simulations rigorously validate the superior performance of the reconstructed model compared to both traditional unconstrained (relaxed) state model and state model without optimized weight matrix. Notably, destination-constrained models exhibit significant advantages not only in capturing the complex dynamics of state evolution under varying start and endpoints but also in generating smoother and more natural trajectories (free from irregular scattering or jumping phenomena), aligning more closely with the actual motion patterns of the target. Via the simulation experiment, we substantiate the tangible benefits of our approach, shedding light on the profound impact that destination constraints wield over the evolution of state trajectories. 

In summary, our work contributes a novel perspective to the field of system modeling in the set-valued framework by seamlessly integrating destination constraints into the fabric of linear dynamic systems. It provides new ideas and important references for improving modeling accuracy and a deeper understanding of the intricate interplay between constraints and system dynamics in the set-valued contexts.
Our next step involves exploring, in the set-valued framework, the extension of linear dynamic systems with destination constraints to nonlinear dynamic systems. This includes considering variations in the types of destination constraints, such as inequalities or other forms, as well as addressing the modeling challenges posed by dynamic systems with multiple waypoints.


\section*{Appendix}
\subsection{Proof of Proposition 1}
We observe that if different values are chosen for the variable $z$ in the relaxed constraint \eqref{eq:13}, the constraints acting on the relaxation constrained combinatorial state $\mathbf{x}_{k}^{rc}$ are different, resulting in state trajectories with destination constraints with different evolutionary trends. Our goal is to find an optimal trajectory for the combination state $\mathbf{x}_{k}^{rc}$ under a specific distance or metric criterion (in this work, we choose the Mahalanobis distance) such that it minimizes the distance to the original constraint combination state $\mathbf{x}_{k}$, while the trajectory of $\mathbf{x}_{k}^{rc}$ must adhere to the original constraints \eqref{eq:12}. This involves solving the optimization problems given by \eqref{eq-15}-\eqref{eq-16}. According to the conditions of Proposition 1, \eqref{eq-15}-\eqref{eq-16} can be rewritten as
\begin{align}\label{eq:15}
	\min_{\mathbf{x}_{k}^{rc}}~~~ &\|\mathbf{x}_{k}^{rc}-\mathbf{x} _{k}\|_{W_{k}}^{2}\\\label{eq:16}
	&s.t. ~~ \mathbf{D}\mathbf{x}_{k}^{rc}=d,
\end{align}
with $W_{k}=\begin{bmatrix}
	W_{k,1} & W_{k,2} \\
	W_{k,2}^{T} & W_{k,3}
\end{bmatrix}  \in \mathbb{R}^{2n\times 2n} $ is a positive definite matrix, $W_{k,\ell}\in \mathbb{R}^{n\times n}, \ell=1,2,3$.
Clearly, \eqref{eq:15}-\eqref{eq:16} form a quadratic programming (QP) problem that can be solved using the method of Lagrange multipliers, resulting in an optimal solution of 
\begin{align}\label{eq:17}
	\mathbf{x}_{k}^{rc}=\mathbf{x}_{k}-W_{k}\mathbf{D}^{T}(\mathbf{D}W_{k}\mathbf{D}^{T})^{-1}(\mathbf{D}\mathbf{x}_{k}-d).  
\end{align}
Based on (\ref{eq:12}), it is easy to see that the right-hand side of (\ref{eq:17}) is equal to $\mathbf{x}_{k} $, i.e.,
\begin{align}\label{eq:18}
	\mathbf{x}_{k}-W_{k}\mathbf{D}^{T}(\mathbf{D}W_{k}\mathbf{D}^{T})^{-1}(\mathbf{D}\mathbf{x}_{k}-d)=\mathbf{x}_{k}.    
\end{align}
\eqref{eq:18} indicates that if the relaxed constraint combination state $\mathbf{x}_{k}^{rc}$ has a solution with the structure of \eqref{eq:17}, then $\mathbf{x}_{k}^{rc}$ will coincide with the original constraint combination state $\mathbf{x}_{k}$. In other words, the Mahalanobis distance between them becomes 0, which is precisely the outcome we desire. Furthermore, considering $\mathbf{D}\mathbf{x}_{k}=d$ in \eqref{eq:12}, then \eqref{eq:18} implies that under the criterion of using Mahalanobis distance as the distance metric, the original constraint combination state $\mathbf{x}_{k}$ has an decomposition form of 
\begin{align}\nonumber
	\mathbf{x}_{k} & =(I-W_{k}\mathbf{D}^{T}(\mathbf{D}W_{k}\mathbf{D}^{T})^{-1}\mathbf{D})\mathbf{x}_{k}\\\nonumber
	&~~~~+W_{k}\mathbf{D}^{T}(\mathbf{D}W_{k}\mathbf{D}^{T})^{-1}\mathbf{D}\mathbf{D}^{\dagger}\mathbf{D}\mathbf{x}_{k}\\\label{eq:19}
	&=A_{k}\mathbf{x}_{k}+(I-A_{k})\mathbf{D}^{\dagger}d,
\end{align}
$A_{k} = I-W_{k}\mathbf{D}^{T}(\mathbf{D}W_{k}\mathbf{D}^{T})^{-1}\mathbf{D} $ and  $\mathbf{D}^{\dagger}=\mathbf{D}^{T}(\mathbf{D}\mathbf{D}^{T})^{-1}$. 
By substituting $W_{k}$ and $\mathbf{D}$ into $A_k$, then
\begin{align}\nonumber
	A_{k}&=I-\begin{bmatrix}
		\mathbf{0}   &  W_{k,2}D^{T}(DW_{k,3}D^{T})^{-1}D\\
		\mathbf{0}  &  W_{k,3}D^{T}(DW_{k,3}D^{T})^{-1}D
	\end{bmatrix}\\\label{eq:20}
	&=\begin{bmatrix}
		I  & -W_{k,2}D^{T}(DW_{k,3}D^{T})^{-1}D\\
		\mathbf{0}  & I-W_{k,3}D^{T}(DW_{k,3}D^{T})^{-1}D
	\end{bmatrix},
\end{align}
Substituting \eqref{eq:20} into \eqref{eq:19} and calculating the first component $x_{k}$ of the combined state $\mathbf{x}_{k}$, we obtain
\begin{align}\nonumber
	x_{k}&=x_{k}-W_{k,2}D^{T}(DW_{k,3}D^{T})^{-1}Dx_{N}\\\label{eq:21}
	&~~~+W_{k,2}D^{T}(DW_{k,3}D^{T})^{-1}d.
\end{align}
Let $t = N$ in (\ref{eq:9}) and substitute it into (\ref{eq:21}) yields
\begin{align}\nonumber
	x_{k} & = x_{k}-W_{k,2}D^{T}(DW_{k,3}D^{T})^{-1}D(\Psi_{k,N}x_{k}+\zeta_{k,N})\\\nonumber
	&~~~+W_{k,2}D^{T}(DW_{k,3}D^{T})^{-1}d\\\nonumber
	&=(I-W_{k,2}D^{T}(DW_{k,3}D^{T})^{-1}D\Psi_{k,N})x_{k}\\\nonumber
	&~~~+W_{k,2}D^{T}(DW_{k,3}D^{T})^{-1}d\\\label{eq:22}
	&~~~-W_{k,2}D^{T}(DW_{k,3}D^{T})^{-1}D\zeta_{k,N}.
\end{align}
Our objective is to establish a recursive state model with destination constraint. Observing \eqref{eq:22}, it can be seen that both states on either side of \eqref{eq:22}, denoted as $x_k$, are subject to destination constraint. Consequently, we can replace the state $x_k$ on the right-hand side of \eqref{eq:22} with the original state model \eqref{eq:4}. Simplifying \eqref{eq:31}, that is,  
\begin{align}\nonumber
\begin{bmatrix}
	\mathbf{0} & D
\end{bmatrix}\begin{bmatrix}
	\mathbf{0} & \mathbf{0}\\
	I & -I
\end{bmatrix}\begin{bmatrix}
	W_{N,1} & W_{N,2}\\
	W_{N,2}^{T} & W_{N,3}
\end{bmatrix}
\begin{bmatrix}
	\mathbf{0} \\
	D^{T}
\end{bmatrix}&=\mathbf{0}\\\label{eq105}
\Longrightarrow
D(W_{N,2}-W_{N,3})D^{T}&=\mathbf{0}.
\end{align}
Combining with \eqref{eq105}, we can achieve the reconstruction of a state model with destination constraint. The established state model with destination-constrained is denoted as 
\begin{align}\nonumber
	x_{k}&=(I-W_{k,2}D^{T}(DW_{k,3}D^{T})^{-1}D\Psi_{k,N})(F_{k-1}x_{k-1}+w_{k-1})\\\nonumber
	&~~~+W_{k,2}D^{T}(DW_{k,3}D^{T})^{-1}d\\\nonumber
	&~~~-W_{k,2}D^{T}(DW_{k,3}D^{T})^{-1}D\zeta_{k,N}\\\nonumber
	&=(I-W_{k,2}D^{T}(DW_{k,3}D^{T})^{-1}D\Psi_{k,N})F_{k-1}x_{k-1}\\\nonumber
	&~~~+(I-W_{k,2}D^{T}(DW_{k,3}D^{T})^{-1}D\Psi_{k,N})w_{k-1}\\\nonumber
	&~~~+W_{k,2}D^{T}(DW_{k,3}D^{T})^{-1}DD^{\dagger}d\\\nonumber
	&~~~-W_{k,2}D^{T}(DW_{k,3}D^{T})^{-1}D\sum_{j=k}^{N-1}\Psi_{j+1,N}w_{j}\\\label{eq:23}
	&=\mathbf{F}_{k-1}x_{k-1}+\mathbf{D}_{k-1}+\mathbf{\Xi}_{k-1}w_{k-1}+\sum_{j=k}^{N-1}\mathbf{\Phi}_{j}w_{j},
\end{align}
with
\begin{align*}
	\mathbf{F}_{k-1}&=\mathbf{\Xi}_{k-1}F_{k-1},\\
	\mathbf{\Xi}_{k-1}&=I-\mathbf{B}_{k-1}\Psi_{k,N}, \\
	\mathbf{B}_{k-1}&=W_{k,2}D^{T}(DW_{k,3}D^{T})^{-1}D, \\
	\mathbf{D}_{k-1}&=\mathbf{B}_{k-1}D^{\dagger}d, \\
	D^{\dagger}&=D^{T}(DD^{T})^{-1}, \\
	\mathbf{\Phi}_{j}&=-\mathbf{B}_{k-1}\Psi_{j+1,N}, j=k,\dots,N-1.
\end{align*}
We have now concluded the proof of Proposition 1.  

\subsection{Proof of Proposition 2}
Our approach is to obtain a feasible solution for \eqref{eq73} by designing $W_{k}$ in \eqref{eq:30} and satisfying \eqref{eq:31}, and to prove that this feasible solution is also the optimal solution. This is accomplished through the following two steps.

\subsubsection{Design of a feasible weighing matrix  to \eqref{eq73}}
It is well known that in a stochastic framework, such as when the process noise is Gaussian white noise, the selection of the weight $W_{k}$ in \eqref{eq:30} based on the covariance matrix of $\mathbf{x}_{k}^{rc}$ reflects the relative uncertainty among its components. By selecting these weights, a state trajectory with destination constraint can be obtained, and the influence of the destination constraint on the state at each time step is relatively uniform. For specific implementation, please refer to  \cite{c30}. However, in the set-valued framework, the distribution of noise is unknown, and the value of the covariance matrix cannot be directly calculated. To the best of our knowledge, there is currently no literature available that specifically addresses the weight matrix $W_{k}$ for the relative uncertainty among components of an unknown but bounded variable $\mathbf{x}_{k}^{rc}$. Therefore, we consider covering the state $\mathbf{x}_{k}^{rc}$ by finding a suitable bounded ellipsoid $\mathcal{E}_{\mathbf{x}_{k}^{rc}}$ in the set-valued framework, quantify the dependent uncertainty between the components of the $\mathbf{x}_{k}^{rc}$ using its shape matrix. The chosen shape matrix of $\mathcal{E}_{\mathbf{x}_{k}^{rc}}$ serves as the weight matrix $W_{k}$ in \eqref{eq:30}. 

Setting $t=N$ in \eqref{eq:9} and replace the original state notation $x_{k}$ with the relaxed state notation $x_{k}^{rc}$, we have 
\begin{align}\nonumber
	x_{N}^{rc}&=\Psi_{k,N}(F_{k-1}x_{k-1}^{rc}+w_{k-1})+\zeta_{k,N}\\\label{eq15}
	&=\Psi_{k-1,N}x_{k-1}^{rc}+\zeta_{k-1,N},
\end{align}
then by the definitions of $\mathbf{x}_{k}^{rc}$ and $\zeta_{k,N}$,
\begin{align}\nonumber
	\mathbf{x}_{k}^{rc}=  \begin{bmatrix}
		x_{k}^{rc} \\
		x_{N}^{rc}
	\end{bmatrix}
	&=\begin{bmatrix}
		F_{k-1}x_{k-1}^{rc}+w_{k-1} \\
		\Psi_{k-1,N}x_{k-1}^{rc}+\sum_{j=k-1}^{N-1}\Psi_{j+1,N}w_{j}
	\end{bmatrix}\\\label{eq17}
	&=\check{\mathbf{x}}_{k}^{rc}+\delta_{k},
\end{align}
where
\begin{align*}
	\check{\mathbf{x}}_{k}^{rc}&=[(F_{k-1}x_{k-1}^{rc})^{T}, (\Psi_{k-1,N}x_{k-1}^{rc})^{T}]^{T},\\
	\delta_{k}&=\mathbf{x}_{k}^{rc}-\check{\mathbf{x}}_{k}^{rc}.
\end{align*}

Observing \eqref{eq17}, we know that when the transition matrix is determined, the uncertainty in updating from state $\mathbf{x}_{k-1}^{rc}$ to $\mathbf{x}_{k}^{rc}$ originates from the process noise. The idea is to use the ellipsoidal $	\mathcal{E}_{\delta_{k}}$ to cover the uncertain part $\delta_{k}$ introduced by process noise. By establishing an affine relationship between the ellipsoidal $	\mathcal{E}_{\delta_{k}}$ and the ellipsoid $\mathcal{W}_{\mathbf{w}_{0}}$, achieving ellipsoidal bounded for the uncertain part $\delta_{k}$. Similarly, through an affine relationship between the ellipsoidal $\mathcal{E}_{\delta_{k}}$ and the ellipsoid $\mathcal{E}_{\mathbf{x}_{k}^{rc}}$ that covers the combined state $\mathbf{x}_{k}^{rc}$ in \eqref{eq17}, we can determine the bounded ellipsoid $\mathcal{E}_{\mathbf{x}_{k}^{rc}}$.
Specifically, we make 
\begin{align}\label{eq47}
	\mathcal{E}_{\mathbf{x}_{k}^{rc}}&=\{\mathbf{x}_{k}^{rc}|(\mathbf{x}_{k}^{rc}-\check{\mathbf{x}}_{k}^{rc})^{T}(\mathbf{P}_{\mathbf{x}_{k}^{rc}})^{-1}(\mathbf{x}_{k}^{rc}-\check{\mathbf{x}}_{k}^{rc})\le 1\},
\end{align} 
where
\begin{align}\label{eq29}
	\check{\mathbf{x}}_{k}^{rc}&=[(F_{k-1}x_{k-1}^{rc})^{T}, (\Psi_{k-1,N}x_{k-1}^{rc})^{T}]^{T},\\\label{eq64}
	\mathbf{P}_{\mathbf{x}_{k}^{rc}} &= \begin{bmatrix}
		P_{k} &  P_{k,N}\\
		P_{N,k} & P_{N}
	\end{bmatrix}.
\end{align} 
It is easy to see that $\mathcal{E}_{\mathbf{x}_{k}^{rc}}$ is an ellipsoidal that covers the composite state $\mathbf{x}_{k}^{rc}$. Its center is at $\check{\mathbf{x}}_{k}^{rc}$, and its shape matrix is $\mathbf{P}_{\mathbf{x}_{k}^{rc}}$. In \eqref{eq64}, $P_k$ and $P_N$ represent the corresponding shape matrices of the ellipsoids containing the states $x_k^{rc}$ and $x_N^{rc}$, respectively. However, $P_{k,N}$ and $P_{N,k}$ can be seen as uncertainty measurement matrices that quantify the relative degree of uncertainty between the states $x_k^{rc}$ and $x_N^{rc}$. Our goal is to obtain an analytical expression for the shape matrix $\mathbf{P}_{\mathbf{x}_{k}^{rc}}$. We consider translating $\mathcal{E}_{\mathbf{x}_{k}^{rc}}$ to have its center at the origin while keeping the shape matrix unchanged. Using $\delta_{k}=\mathbf{x}_{k}^{rc}-\check{\mathbf{x}}_{k}^{rc}$, we denote the ellipsoidal after this translation as 
\begin{align}\nonumber
	\mathcal{E}_{\delta_{k}}&:=\mathcal{E}_{\delta_{k}}(\mathbf{c}_{\delta_{k}},\mathbf{P}_{\delta_{k}})=\mathcal{E}_{\delta_{k}}(\mathbf{0},\mathbf{P}_{\mathbf{x}_{k}^{rc}})\\\label{eq62}
	&=\{\delta_{k}|(\delta_{k}-\mathbf{0})^{T}(\mathbf{P}_{\mathbf{x}_{k}^{rc}})^{-1}(\delta_{k}-\mathbf{0})\le 1\}.
\end{align} 
It is worth noting that
\begin{align}\nonumber
	\mathbf{x}_{k}^{rc}-\check{\mathbf{x}}_{k}^{rc}&=\begin{bmatrix}
		w_{k-1} \\
		\sum_{j=k-1}^{N-1}\Psi_{j+1,N}w_{j}
	\end{bmatrix}\\\nonumber
	&=\begin{bmatrix}
		I & \mathbf{0} & \cdots & \mathbf{0}\\
		\Psi_{k,N}  & \Psi_{k+1,N} & \cdots & \Psi_{N,N}
	\end{bmatrix}
	\begin{bmatrix}
		w_{k-1} \\
		w_{k} \\
		\vdots  \\
		w_{N-1}
	\end{bmatrix}\\\label{eq18}
	&=\mathbf{C}_{k}\mathbf{w}_{k-1},
\end{align}
where $\mathbf{C}_{k}=[\mathbf{G}_{k}^{T}, \mathbf{\Psi}_{k}^{T}]^{T}$, the definitions of $\mathbf{G}_{k}$ and $\mathbf{\Psi}_{k}$ can be viewed \eqref{eq16}. By (\ref{eq59}), (\ref{eq18}), and the definition of $\delta_{k}$, we have 
\begin{align}\label{eq61}
	\delta_{k}=\mathbf{C}_{k}M_{k}\mathbf{w}_{0},
\end{align}
the definition of $M_{k}$ is given in (\ref{eq60}).
In (\ref{eq61}), matrices $\mathbf{C}_{k}$ and $M_{k}$ are known, while the vector $\mathbf{w}_{0}$ is uncertain. According to the conditions of Proposition 2, we know that $\mathbf{w}_{0} \in\mathcal{W}_{\mathbf{w}_{0}}$. From (\ref{eq61}), the $\delta_{k}$ representation is obtained by affine transformation of the vector $\mathbf{w}_{0}$. Utilizing the properties of elliptic affine transformations, as detailed in \cite{c50}. Therefore, there is the following relationship between the ellipsoid $\mathcal{E}_{\delta_{k}}$ covering the state $\delta_{k}$ and the ellipsoid $\mathcal{W}_{\mathbf{w}_{0}}$ covering the combined process noise vector $\mathbf{w}_{0}$: 
\begin{align}\label{eq30}
	\mathcal{E}_{\delta_{k}}=\mathbf{C}_{k}M_{k}\mathcal{W}_{\mathbf{w}_{0}},
\end{align}
the center and shape matrix corresponding to the ellipsoid satisfy the following relationships:
\begin{align}\label{eq20}
	\mathbf{c}_{\delta_{k}}&=\mathbf{0}=\mathbf{C}_{k}M_{k}\cdot\mathbf{0},\\\label{eq46}
	\mathbf{P}_{\delta_{k}}&=\mathbf{P}_{\mathbf{x}_{k}^{rc}}=\mathbf{C}_{k}M_{k}Q_{\mathbf{w}_{0}}(\mathbf{C}_{k}M_{k})^{T}.
\end{align}
Taking into account (\ref{eq62}), it is easy to generate 
\begin{align}\nonumber
	\mathcal{E}_{\delta_{k}}&=
	\{\delta_{k}|\delta_{k}^{T}(\mathbf{P}_{\mathbf{x}_{k}^{rc}})^{-1}\delta_{k}\le 1\}\\\nonumber
	&=\{\delta_{k}|(\delta_{k}-\mathbf{0})^{T}(\mathbf{C}_{k}M_{k}Q_{\mathbf{w}_{0}}(\mathbf{C}_{k}M_{k})^{T})^{-1}(\delta_{k}-\mathbf{0})\le 1\}
\end{align}
This indicates that although ellipsoids $\mathcal{E}_{\delta_{k}}$ and $\mathcal{E}_{\mathbf{x}_{k}^{rc}}$ have different centers, they share the same shape matrix $\mathbf{P}_{\mathbf{x}_{k}^{rc}}$. 
Based on \eqref{eq64}, the shape matrix $\mathbf{P}_{\mathbf{x}_{k}^{rc}}$ of  $\mathcal{E}_{\mathbf{x}_{k}^{rc}}$ can reflect the relative uncertainty among the components of $\mathbf{x}_{k}^{rc}$.
In a manner analogous to the selection of the covariance matrix of a random vector as a weight matrix in the random framework, we choose the shape matrix $\mathbf{P}_{\mathbf{x}_{k}^{rc}}$ of $\mathcal{E}_{\mathbf{x}_{k}^{rc}}$ as the weight matrix $W_{k}$ in \eqref{eq:30} under the set-valued framework, which can achieve a similar weighted effect and satisfies 
\begin{align}\label{eq22}
	W_{k}=\mathbf{P}_{\mathbf{x}_{k}^{rc}}=\mathbf{C}_{k}M_{k}Q_{\mathbf{w}_{0}}M_{k}^{T}\mathbf{C}_{k}^{T}.
\end{align}
By substituting \eqref{eq16}, \eqref{eq60}, \eqref{eq23}, and \eqref{eq85} into \eqref{eq22} and leveraging the properties of matrix operations \cite{c51}, we have
\begin{align}\nonumber
	W_{k}&=\mathbf{C}_{k}(M_{k}Q_{\mathbf{w}_{0}}M_{k}^{T})\mathbf{C}_{k}^{T}\\\nonumber
	&=\mathbf{C}_{k}Q_{\mathbf{w}_{k-1}}\mathbf{C}_{k}^{T}\\\nonumber
	&=\begin{bmatrix}
		I & \mathbf{0} & \cdots & \mathbf{0}\\
		\Psi_{k,N}  & \Psi_{k+1,N} & \cdots & \Psi_{N,N}
	\end{bmatrix}\\\nonumber
	&~~~\begin{bmatrix}
		Q_{k-1} & Q_{k-1,k} & \cdots & Q_{k-1,N-1}\\
		Q_{k,k-1} & Q_{k} & \cdots & Q_{k,N-1}\\
		\vdots & \vdots & \vdots & \vdots\\
		Q_{N-1,k-1} & Q_{N-1,k} & \cdots & Q_{N-1}
	\end{bmatrix}
	\begin{bmatrix}
		I & \Psi_{k,N}^{T}\\
		\mathbf{0} & \Psi_{k+1,N}^{T}\\
		\vdots  & \vdots\\
		\mathbf{0} & \Psi_{N,N}^{T}
	\end{bmatrix}\\\label{eq24}
	&=\begin{bmatrix}
		Q_{k-1} & \mathbf{G}_{k}Q_{\mathbf{w}_{k-1}}\mathbf{\Psi}_{k}^{T} \\
		\mathbf{\Psi}_{k}Q_{\mathbf{w}_{k-1}}\mathbf{G}_{k}^{T} & \mathbf{\Psi}_{k}Q_{\mathbf{w}_{k-1}}\mathbf{\Psi}_{k}^{T}
	\end{bmatrix}.
\end{align}
Based on \eqref{eq24}, \eqref{eq85}, \eqref{eq60}, and \eqref{eq23}, we obtain the analytical expression for the weight matrix $W_{k}$ in \eqref{eq:30}, where
\begin{align}\label{eq108}
	W_{k,1}&=Q_{k-1}\\\nonumber
	W_{k,2}&=\mathbf{G}_{k}Q_{\mathbf{w}_{k-1}}\mathbf{\Psi}_{k}^{T}=\mathbf{G}_{k}M_{k}Q_{\mathbf{w}_{0}}M_{k}^{T}\mathbf{\Psi}_{k}^{T}\\\label{eq27}
	&=\sum_{i=k-1}^{N-1}Q_{k-1,i}\Psi_{i+1,N}^{T},\\\nonumber
	W_{k,3}&=\mathbf{\Psi}_{k}Q_{\mathbf{w}_{k-1}}\mathbf{\Psi}_{k}^{T}=\mathbf{\Psi}_{k}M_{k}Q_{\mathbf{w}_{0}}M_{k}^{T}\mathbf{\Psi}_{k}^{T}\\\label{eq28} 
	&=\sum_{j=k-1}^{N-1}
	\sum_{i=k-1}^{N-1}\Psi_{i+1,N}Q_{i,j}\Psi_{j+1,N}^{T}.
\end{align}
Obviously, the designed $W_{k}\succ \mathbf{0}$ (as it represents the shape matrix of the ellipsoid $\mathcal{E}_{\mathbf{x}_{k}^{rc}}$) in \eqref{eq24} and $D(W_{N,2}-W_{N,3})D^{T}=\mathbf{0}$ satisfy the model \eqref{eq:26}. It can be verified that it is a feasible solution to \eqref{eq73}.

\subsubsection{Optimality of the Weight Matrix}
For model \eqref{eq:26}, our goal is to minimize the shape matrix of the ellipsoid covering the process noise generated during the update from the state $x_{k-1}$ to $x_{k}$. This has been formulated as an optimization problem \eqref{eq73}. Observing the objective function and optimization variables of \eqref{eq73}, based on the definitions of $Q_{\eta_{k}}(W_{k}))$, $H_{k}$, and $\mathbf{B}_{k-1}$, the optimization problem \eqref{eq73} can be equivalently reformulated as the following optimization problem:
\begin{align}\label{eq104}
	\min_{W_{k,2},W_{k,3}}& ~~ Q_{\eta_{k}}(W_{k})\\\nonumber
	\text{s.t.}~~& W_{k,1} \succ \mathbf{0}, W_{k,3} \succ \mathbf{0},\\\nonumber
	&W_{k,1}-W_{k,2}W_{k,3}^{-1}W_{k,2}^{T} \succ \mathbf{0},
\end{align}
Considering that $W_{k}$ in \eqref{eq24} is a feasible solution from \eqref{eq73}, evidently, $W_{k,2}$ in \eqref{eq27} and  $W_{k,3}$ \eqref{eq28} are feasible solution for \eqref{eq104}.  
Next, let $W_{k}^{*}=W_{k}$, where $W_{k}$ is defined in \eqref{eq24}, and we can show that $W_{k,2}^{*}$ and $W_{k,3}^{*}$ are optimal solutions for \eqref{eq104}. To avoid confusion of symbols, we use $\tilde{W}_{k}=\begin{bmatrix}
	\tilde{W}_{k,1} & \tilde{W}_{k,2} \\
	\tilde{W}_{k,2}^{T} & \tilde{W}_{k,3}
\end{bmatrix} \ne W_{k}^{*}$ to denote any weight matrix that satisfies the constraints in optimization problem \eqref{eq104}.  
Due to \eqref{eq:26}, \eqref{eq16}, \eqref{eq84}, \eqref{eq27}, and \eqref{eq28}, we have the following notation:
\begin{align*}
	Q_{\eta_{k}}(W_{k}^{*})&=H_{k}^{*}Q_{\mathbf{w}_{k-1}}(H_{k}^{*})^{T}\\ Q_{\eta_{k}}(\tilde{W}_{k})&=\tilde{H}_{k}Q_{\mathbf{w}_{k-1}}\tilde{H}_{k}^{T}\\
	H_{k}^{*}&=\mathbf{G}_{k}-\mathbf{B}_{k-1}^{*}\mathbf{\Psi}_{k},\\
	\tilde{H}_{k}&=\mathbf{G}_{k}-\tilde{\mathbf{B}}_{k-1}\mathbf{\Psi}_{k},\\
	\mathbf{B}_{k-1}^{*}&=W_{k,2}^{*}D^{T}(DW_{k,3}^{*}D^{T})^{-1}D,\\
	\tilde{\mathbf{B}}_{k-1}&=\tilde{W}_{k,2}D^{T}(D\tilde{W}_{k,3}D^{T})^{-1}D,\\
	W_{k,2}^{*}&=\mathbf{G}_{k}Q_{\mathbf{w}_{k-1}}\mathbf{\Psi}_{k},\\
	W_{k,3}^{*}&=\mathbf{\Psi}_{k}Q_{\mathbf{w}_{k-1}}\mathbf{\Psi}_{k}^{T},
\end{align*}
where $\tilde{W}_{k,2}$ and $\tilde{W}_{k,3}\succ \mathbf{0}$ are arbitrary matrices with appropriate dimensions. From \eqref{eq63}, $Q_{\mathbf{w}_{0}}$ is the shape matrix of the ellipsoid $\mathcal{W}_{\mathbf{w}_{0}}$ is also a positive definite matrix. Based on \eqref{eq85}, utilizing the properties of positive definite matrices \cite{c52}, we can derive that $Q_{\mathbf{w}_{k-1}}$ is also a positive definite matrix, denoted as $Q_{\mathbf{w}_{k-1}}\succ \mathbf{0}$. Similarly, for the matrix $\tilde{H}_{k}-H_{k}^{*}$, using the properties of positive definite matrices, we can find that 
\begin{align}\nonumber
	&(\tilde{H}_{k}-H_{k}^{*})Q_{\mathbf{w}_{k-1}}(\tilde{H}_{k}-H_{k}^{*})^{T}\\\nonumber
	&=\tilde{H}_{k}Q_{\mathbf{w}_{k-1}}\tilde{H}_{k}^{T}-\tilde{H}_{k}Q_{\mathbf{w}_{k-1}}(H_{k}^{*})^{T}-H_{k}^{*}Q_{\mathbf{w}_{k-1}}\tilde{H}_{k}^{T}\\\nonumber
	&~~~+H_{k}^{*}Q_{\mathbf{w}_{k-1}}(H_{k}^{*})^{T}\\\label{eq77}
	&\succeq \mathbf{0}.
\end{align}
Actually, we can compute the analytic expressions for each of the three terms $\tilde{H}_{k}Q_{\mathbf{w}_{k-1}}(H_{k}^{*})^{T}$, $H_{k}^{*}Q_{\mathbf{w}_{k-1}}\tilde{H}_{k}^{T}$, and $H_{k}^{*}Q_{\mathbf{w}_{k-1}}(H_{k}^{*})^{T}$ in \eqref{eq77} and deduce that they are equivalent. 
\begin{align}\nonumber
	&\tilde{H}_{k}Q_{\mathbf{w}_{k-1}}(H_{k}^{*})^{T}\\\nonumber
	&=(\mathbf{G}_{k}-\tilde{\mathbf{B}}_{k-1}\mathbf{\Psi}_{k})Q_{\mathbf{w}_{k-1}}(\mathbf{G}_{k}-\mathbf{B}_{k-1}^{*}\mathbf{\Psi}_{k})^{T}\\\nonumber
	&=\mathbf{G}_{k}Q_{\mathbf{w}_{k-1}}\mathbf{G}_{k}^{T}-\mathbf{G}_{k}Q_{\mathbf{w}_{k-1}}\mathbf{\Psi}_{k}^{T}(\mathbf{B}_{k-1}^{*})^{T}\\\nonumber
	&~~~-\tilde{\mathbf{B}}_{k-1}\mathbf{\Psi}_{k}Q_{\mathbf{w}_{k-1}}\mathbf{G}_{k}^{T}+\tilde{\mathbf{B}}_{k-1}\mathbf{\Psi}_{k}Q_{\mathbf{w}_{k-1}}\mathbf{\Psi}_{k}^{T}(\mathbf{B}_{k-1}^{*})^{T}\\\nonumber
	&=\mathbf{G}_{k}Q_{\mathbf{w}_{k-1}}\mathbf{G}_{k}^{T}-W_{k,2}^{*}(\mathbf{B}_{k-1}^{*})^{T}-\tilde{\mathbf{B}}_{k-1}(W_{k,2}^{*})^{T}\\\nonumber
	&~~~+\tilde{\mathbf{B}}_{k-1}W_{k,3}^{*}(\mathbf{B}_{k-1}^{*})^{T}\\\nonumber
	&=\mathbf{G}_{k}Q_{\mathbf{w}_{k-1}}\mathbf{G}_{k}^{T}-W_{k,2}^{*}D^{T}(DW_{k,3}^{*}D^{T})^{-1}D(W_{k,2}^{*})^{T}\\\nonumber
	&~~~-\tilde{W}_{k,2}D^{T}(D\tilde{W}_{k,3}D^{T})^{-1}D(W_{k,2}^{*})^{T}\\\nonumber
	&~~~+\tilde{W}_{k,2}D^{T}(D\tilde{W}_{k,3}D^{T})^{-1}D(W_{k,2}^{*})^{T}\\\label{eq79}
	&=\mathbf{G}_{k}Q_{\mathbf{w}_{k-1}}\mathbf{G}_{k}^{T}-W_{k,2}^{*}D^{T}(DW_{k,3}^{*}D^{T})^{-1}D(W_{k,2}^{*})^{T},
\end{align}
\begin{align}\nonumber
	&H_{k}^{*}Q_{\mathbf{w}_{k-1}}\tilde{H}_{k}^{T}\\\nonumber
	&=(\mathbf{G}_{k}-\mathbf{B}_{k-1}^{*}\mathbf{\Psi}_{k})Q_{\mathbf{w}_{k-1}}(\mathbf{G}_{k}-\tilde{\mathbf{B}}_{k-1}\mathbf{\Psi}_{k})^{T}\\\nonumber
	&=\mathbf{G}_{k}Q_{\mathbf{w}_{k-1}}\mathbf{G}_{k}^{T}-\mathbf{G}_{k}Q_{\mathbf{w}_{k-1}}\mathbf{\Psi}_{k}^{T}\tilde{\mathbf{B}}_{k-1}^{T}\\\nonumber
	&~~~-\mathbf{B}_{k-1}^{*}\mathbf{\Psi}_{k}Q_{\mathbf{w}_{k-1}}\mathbf{G}_{k}^{T}+\mathbf{B}_{k-1}^{*}\mathbf{\Psi}_{k}Q_{\mathbf{w}_{k-1}}\mathbf{\Psi}_{k}^{T}\tilde{\mathbf{B}}_{k-1}^{T}\\\nonumber
	&=\mathbf{G}_{k}Q_{\mathbf{w}_{k-1}}\mathbf{G}_{k}^{T}-W_{k,2}^{*}\tilde{\mathbf{B}}_{k-1}^{T}-\mathbf{B}_{k-1}^{*}(W_{k,2}^{*})^{T}+\mathbf{B}_{k-1}^{*}W_{k,3}^{*}\tilde{\mathbf{B}}_{k-1}^{T}\\\nonumber
	&=\mathbf{G}_{k}Q_{\mathbf{w}_{k-1}}\mathbf{G}_{k}^{T}-W_{k,2}^{*}D^{T}(DW_{k,3}^{*}D^{T})^{-1}D\tilde{W}_{k,2}^{T}\\\nonumber
	&~~~-W_{k,2}^{*}D^{T}(DW_{k,3}^{*}D^{T})^{-1}D(W_{k,2}^{*})^{T}\\\nonumber
	&~~~+W_{k,2}^{*}D^{T}(DW_{k,3}^{*}D^{T})^{-1}D\tilde{W}_{k,2}^{T}\\\label{eq80}
	&=\mathbf{G}_{k}Q_{\mathbf{w}_{k-1}}\mathbf{G}_{k}^{T}-W_{k,2}^{*}D^{T}(DW_{k,3}^{*}D^{T})^{-1}D(W_{k,2}^{*})^{T},
\end{align}
\begin{align}\nonumber
	&H_{k}^{*}Q_{\mathbf{w}_{k-1}}(H_{k}^{*})^{T}\\\nonumber
	&=(\mathbf{G}_{k}-\mathbf{B}_{k-1}^{*}\mathbf{\Psi}_{k})Q_{\mathbf{w}_{k-1}}(\mathbf{G}_{k}-\mathbf{B}_{k-1}^{*}\mathbf{\Psi}_{k})^{T}\\\nonumber
	&=\mathbf{G}_{k}Q_{\mathbf{w}_{k-1}}\mathbf{G}_{k}^{T}-\mathbf{G}_{k}Q_{\mathbf{w}_{k-1}}\mathbf{\Psi}_{k}^{T}(\mathbf{B}_{k-1}^{*})^{T}\\\nonumber
	&~~~-\mathbf{B}_{k-1}^{*}\mathbf{\Psi}_{k}Q_{\mathbf{w}_{k-1}}\mathbf{G}_{k}^{T}+\mathbf{B}_{k-1}^{*}\mathbf{\Psi}_{k}Q_{\mathbf{w}_{k-1}}\mathbf{\Psi}_{k}^{T}(\mathbf{B}_{k-1}^{*})^{T}\\\nonumber
	&=\mathbf{G}_{k}Q_{\mathbf{w}_{k-1}}\mathbf{G}_{k}^{T}-W_{k,2}^{*}(\mathbf{B}_{k-1}^{*})^{T}-\mathbf{B}_{k-1}^{*}(W_{k,2}^{*})^{T}\\\nonumber
	&~~~+\mathbf{B}_{k-1}^{*}W_{k,3}^{*}(\mathbf{B}_{k-1}^{*})^{T}\\\nonumber
	&=\mathbf{G}_{k}Q_{\mathbf{w}_{k-1}}\mathbf{G}_{k}^{T}-W_{k,2}^{*}D^{T}(DW_{k,3}^{*}D^{T})^{-1}D(W_{k,2}^{*})^{T}\\\nonumber
	&~~~-W_{k,2}^{*}D^{T}(DW_{k,3}^{*}D^{T})^{-1}D(W_{k,2}^{*})^{T}\\\nonumber
	&~~~+W_{k,2}^{*}D^{T}(DW_{k,3}^{*}D^{T})^{-1}D(W_{k,2}^{*})^{T}\\\label{eq81}
	&=\mathbf{G}_{k}Q_{\mathbf{w}_{k-1}}\mathbf{G}_{k}^{T}-W_{k,2}^{*}D^{T}(DW_{k,3}^{*}D^{T})^{-1}D(W_{k,2}^{*})^{T}.
\end{align}
Substituting \eqref{eq79}, \eqref{eq80}, and \eqref{eq81} into \eqref{eq77} yields 
\begin{align}\nonumber
	&\tilde{H}_{k}Q_{\mathbf{w}_{k-1}}\tilde{H}_{k}^{T}-\tilde{H}_{k}Q_{\mathbf{w}_{k-1}}(H_{k}^{*})^{T}-H_{k}^{*}Q_{\mathbf{w}_{k-1}}\tilde{H}_{k}^{T}\\\nonumber
	&~~~+H_{k}^{*}Q_{\mathbf{w}_{k-1}}(H_{k}^{*})^{T}\\\label{eq82}
	&=\tilde{H}_{k}Q_{\mathbf{w}_{k-1}}\tilde{H}_{k}^{T}-H_{k}^{*}Q_{\mathbf{w}_{k-1}}(H_{k}^{*})^{T}\succeq  \mathbf{0}.
\end{align}
By the definitions of $Q_{\eta_{k}}(W_{k}^{*})$ and $Q_{\eta_{k}}(\tilde{W}_{k})$, together with \eqref{eq82}, we can deduce that 
\begin{align}\label{eq107}
	Q_{\eta_{k}}(W_{k}^{*})=H_{k}^{*}Q_{\mathbf{w}_{k-1}}(H_{k}^{*})^{T}\preceq \tilde{H}_{k}Q_{\mathbf{w}_{k-1}}\tilde{H}_{k}^{T}=Q_{\eta_{k}}(\tilde{W}_{k}).
\end{align}

According to \eqref{eq107}, exploiting the arbitrariness of the matrix $\tilde{H}_{k}$, which is determined by arbitrary weight matrix $\tilde{W}_{k} \succ \mathbf{0}$, it is evident that we have proven the optimality of the sub-block matrices $W_{k,2}^{*}$ and $W_{k,3}^{*}$ for the optimization problem \eqref{eq104}. Building upon the definition of $W_{k}^{*}$ in Proposition 2 and considering the equivalence relationship between \eqref{eq73} and \eqref{eq104}, it is evident that we can deduce that $W_{k}^{*}$ in \eqref{eq111} is an optimal solution to \eqref{eq73}. Additionally, the value of $W_{k,1}^{*}$ in \eqref{eq112} is not unique due to the fact that $W_{k,1}^{*}$ is not a decision variable in \eqref{eq73}, and therefore, it does not affect the objective function's value. Hence, the value of $W_{k,1}^{*}=W_{k,1}^{**}$ only needs to ensure that the weight matrix $W_{k}^{*}$ is positive definite, i.e., $W_{k,1}^{**}\succ \mathbf{0}$ and $W_{k,1}^{**}-W_{k,2}^{*}(W_{k,3}^{*})^{-1}(W_{k,2}^{*})^{T}\succ \mathbf{0}$. Simultaneously, selecting \eqref{eq109} and \eqref{eq113} as sub-block matrices $W_{k,2}^{*}$ and $W_{k,3}^{*}$ and substituting them into \eqref{eq73} yields the same optimal values, indicating that the constructed $W_{k}^{*}$ is also an optimal solution to \eqref{eq73}. So far, we have completed the proof of Proposition 2.  

\subsection{Proof of Proposition 3}
Based on model \eqref{eq:4}, we know that when the state transfer matrix $F_{k-1}$ is given, the uncertainty in updating the state from $x_{k-1}$ to $x_{k}$ due to the process noise can be expressed as \begin{align}\label{eq92}
	w_{k-1}=x_{k}-F_{k-1}x_{k-1}.
\end{align}
By \eqref{eq:6}, the noise part $w_{k-1}$ belongs to a known bounded ellipsoid $\mathcal{W}_{k-1}=\mathcal{E}_{w_{k-1}}(\mathbf{0},Q_{k-1})$. Similarly, by employing \eqref{eq74}, the process noise generated during state updates for the model given by \eqref{eq:26} can be expressed as:
\begin{align}\label{eq93}
	\eta_{k}=H_{k}M_{k}\mathbf{w}_{0}.
\end{align}
From the proof procedure of Proposition 2, combined with \eqref{eq76} and \eqref{eq84}, we know that the vector $\eta_{k}$ belongs to the bounded ellipsoid  
\begin{align}\label{eq94}
	\mathcal{W}_{\eta_{k}}=\mathcal{W}_{\eta_{k}}(\mathbf{c}_{\eta_{k}},Q_{\eta_{k}})=\mathcal{W}_{\eta_{k}}(\mathbf{0},H_{k}Q_{\mathbf{w}_{k-1}}H_{k}^{T}).  
\end{align}
When model \eqref{eq:26} is configured with the optimal weight matrix $W_{k}^{*}$, the shape matrix of the ellipsoid covering $\eta_{k}$ becomes $Q_{\eta_{k}}(W_{k}^{*})=H_{k}^{*}Q_{\mathbf{w}_{k-1}}(H_{k}^{*})^{T}$. 
Based on \eqref{eq16},  \eqref{eq60}, \eqref{eq23}, \eqref{eq85}, and \eqref{eq81}, then we have 
\begin{align}\nonumber
	Q_{\eta_{k}}(W_{k}^{*})&=\mathbf{G}_{k}Q_{\mathbf{w}_{k-1}}\mathbf{G}_{k}^{T}-W_{k,2}^{*}D^{T}(DW_{k,3}^{*}D^{T})^{-1}D(W_{k,2}^{*})^{T}\\\label{eq95}
	&=Q_{k-1}-W_{k,2}^{*}D^{T}(DW_{k,3}^{*}D^{T})^{-1}D(W_{k,2}^{*})^{T}.
\end{align}
Since the weight matrix $W_{k,3}^{*}\succ \mathbf{0}$ is positive definite, it is easy to verify that $W_{k,2}^{*}D^{T}(DW_{k,3}^{*}D^{T})^{-1}D(W_{k,2}^{*})^{T}\succeq \mathbf{0}$ using the properties of positive definite matrices. Thus, for \eqref{eq95}, we can deduce that \begin{align}\nonumber
	Q_{\eta_{k}}(W_{k}^{*})&=Q_{k-1}-W_{k,2}^{*}D^{T}(DW_{k,3}^{*}D^{T})^{-1}D(W_{k,2}^{*})^{T}\\\label{eq96}
	&\preceq  Q_{k-1}.
\end{align}
This completes the proof of Proposition 3.

\subsection{Proof of Proposition 4}
Under the conditions of Proposition 4, the state model \eqref{eq:4} can be reformulated as 
\begin{align}\label{eq98}
	x_{k}&=Fx_{k-1}+w_{k-1}, w_{k-1}\in \mathcal{W}=\mathcal{E}_{w_{k-1}}(0,Q).
\end{align}
Likewise, by \eqref{eq109} and \eqref{eq113}, the optimal sub-block matrices used in model \eqref{eq:26} become 
\begin{align}\label{eq99}
	W_{k,2}^{*}&=\sum_{i=k-1}^{N-1}Q_{k-1,i}\Psi_{i+1,N}^{T}=Q(F^{N-k})^{T},\\\nonumber
	W_{k,3}^{*}&=\sum_{j=k-1}^{N-1}
	\sum_{i=k-1}^{N-1}\Psi_{i+1,N}Q_{i,j}\Psi_{j+1,N}^{T}\\\label{eq100}
	&=\sum_{i=k-1}^{N-1}F^{N-i-1}Q(F^{N-i-1})^{T}.
\end{align}
Because the matrices $F$ and $D$ are both invertible, based on \eqref{eq95}, we can derive that
\begin{align}\nonumber
	&Q_{\eta_{k}}(W_{k}^{*})\\\nonumber
	&=H_{k}^{*}Q_{\mathbf{w}_{k-1}}(H_{k}^{*})^{T}\\\nonumber
	&=Q_{k-1}-W_{k,2}^{*}D^{T}(DW_{k,3}^{*}D^{T})^{-1}D(W_{k,2}^{*})^{T}\\\nonumber
	&=Q-Q(F^{N-k})^{T}D^{T}(D\sum_{i=k-1}^{N-1}F^{N-i-1}Q\\\nonumber
	&~~~(F^{N-i-1})^{T}D^{T})^{-1}D(Q(F^{N-k})^{T})^{T}\\\label{eq101}
	&=Q-Q(F^{N-k})^{T}(\sum_{i=k}^{N}F^{N-i}Q(F^{N-i})^{T})^{-1}(Q(F^{N-k})^{T})^{T},
\end{align}
\begin{align}\nonumber
	&Q_{\eta_{k+1}}(W_{k+1}^{*})\\\nonumber
	&=H_{k+1}^{*}Q_{\mathbf{w}_{k}}(H_{k+1}^{*})^{T}\\\nonumber
	&=Q_{k}-W_{k+1,2}^{*}D^{T}(DW_{k+1,3}^{*}D^{T})^{-1}D(W_{k+1,2}^{*})^{T}\\\nonumber
	&=Q-Q(F^{N-k-1})^{T}D^{T}(D\sum_{i=k}^{N-1}F^{N-i-1}Q\\\nonumber
	&~~~(F^{N-i-1})^{T}D^{T})^{-1}D(Q(F^{N-k-1})^{T})^{T}\\\nonumber
	&=Q-Q(F^{N-k-1})^{T}(\sum_{i=k}^{N-1}F^{N-i-1}Q\\\nonumber
	&~~~(F^{N-i-1})^{T})^{-1}(Q(F^{N-k-1})^{T})^{T}\\\nonumber
	&=Q-Q(F^{N-k-1})^{T}(F^{-1}F\sum_{i=k}^{N-1}F^{N-i-1}Q\\\nonumber
	&~~~(F^{N-i-1})^{T}F^{T}(F^{T})^{-1})^{-1}(Q(F^{N-k-1})^{T})^{T}\\\label{eq102}
	&=Q-Q(F^{N-k})^{T}(\sum_{i=k}^{N-1}F^{N-i}Q(F^{N-i})^{T})^{-1}(Q(F^{N-k})^{T})^{T}.
\end{align}
Using \eqref{eq101} minus \eqref{eq102}, we have
\begin{align}\nonumber
	&Q_{\eta_{k}}(W_{k}^{*})-Q_{\eta_{k+1}}(W_{k+1}^{*})\\\nonumber
	&=Q(F^{N-k})^{T}(\sum_{i=k}^{N-1}F^{N-i}Q(F^{N-i})^{T})^{-1}(Q(F^{N-k})^{T})^{T}\\\nonumber
	&~~~-Q(F^{N-k})^{T}(\sum_{i=k}^{N}F^{N-i}Q(F^{N-i})^{T})^{-1}(Q(F^{N-k})^{T})^{T}\\\nonumber
	&=Q(F^{N-k})^{T}[(\sum_{i=k}^{N-1}F^{N-i}Q(F^{N-i})^{T})^{-1}\\\label{eq103}
	&~~~-(\sum_{i=k}^{N}F^{N-i}Q(F^{N-i})^{T})^{-1}](Q(F^{N-k})^{T})^{T}.
\end{align}
Since $Q$ is a positive definite matrix, then 
\begin{align}\label{eq114}
	\sum_{i = k}^{N}F^{N-i}Q(F^{N-i})^{T}-\sum_{i = k}^{N-1}F^{N-i}Q(F^{N-i})^{T} = Q\succ \mathbf{0}.
\end{align}
By \eqref{eq114}, we obtain  
\begin{align}\nonumber
	\sum_{i = k}^{N}F^{N-i}Q(F^{N-i})^{T}&\succ\sum_{i = k}^{N-1}F^{N-i}Q(F^{N-i})^{T}\\\label{eq115}
	\Longrightarrow (\sum_{i = k}^{N}F^{N-i}Q(F^{N-i})^{T})^{-1}&\prec (\sum_{i = k}^{N-1}F^{N-i}Q(F^{N-i})^{T})^{-1}.
\end{align}
Combining \eqref{eq103} and \eqref{eq115}, we can derive 
\begin{align}\label{eq116}
	Q_{\eta_{k}}(W_{k}^{*})-Q_{\eta_{k+1}}(W_{k+1}^{*})\succeq \mathbf{0}.
\end{align}
According to \eqref{eq116}, we have that \eqref{eq97} holds. Up to this point, the proof of Proposition 4 has been concluded.

\section*{References}

\def\refname{\vadjust{\vspace*{-2.5em}}}

\end{document}